\documentclass[10pt,twocolumn,preprintnumbers,prd,superscriptaddress,nofootinbib,floatfix,showpacs,showkeys,aps]{revtex4-1}

\usepackage{amsmath}
\usepackage{amsfonts}
\usepackage{amssymb}
\usepackage{graphicx}
\usepackage[titletoc]{appendix}
\usepackage{color}
\usepackage{hyperref}
\usepackage{cleveref}
\usepackage[rightcaption]{sidecap}
\usepackage{subcaption}
\usepackage{comment}

\usepackage{mathtools}

\usepackage{dcolumn}

\usepackage{array}
\usepackage{ctable}
\usepackage{multirow}
\usepackage{siunitx}
\usepackage{tabularx}
\usepackage{booktabs}
\usepackage{supertabular}
\usepackage{verbatim}

\usepackage{ragged2e}
\usepackage{caption}
\DeclareCaptionJustification{justified}{\justifying}
\usepackage{cancel}
\usepackage{graphicx}
\usepackage{dcolumn}
\usepackage{bm}
\usepackage{braket}

\begin{document}

\title{Entanglement from quantum massless spin-2 perturbations during inflation}

\author{Mattia Dubbini}
\email{mattia.dubbini@unicam.it}
\affiliation{University of Camerino, Via Madonna delle Carceri, Camerino, 62032, Italy.}
\affiliation{Istituto Nazionale di Fisica Nucleare (INFN), Sezione di Perugia, Perugia, 06123, Italy.}

\author{Orlando Luongo}
\email{orlando.luongo@unicam.it}
\affiliation{University of Camerino, Via Madonna delle Carceri, Camerino, 62032, Italy.}
\affiliation{Istituto Nazionale di Fisica Nucleare (INFN), Sezione di Perugia, Perugia, 06123, Italy.}

\affiliation{Department of Nanoscale Science and Engineering,
University at Albany-SUNY, Albany, New York 12222, USA.}
\affiliation{INAF - Osservatorio Astronomico di Brera, Milano, Italy.}
\affiliation{Al-Farabi Kazakh National University, Almaty, 050040, Kazakhstan.}

\author{Stefano Mancini}
\email{stefano.mancini@unicam.it}
\affiliation{University of Camerino, Via Madonna delle Carceri, Camerino, 62032, Italy.}
\affiliation{Istituto Nazionale di Fisica Nucleare (INFN), Sezione di Perugia, Perugia, 06123, Italy.}

\author{Tommaso Mengoni}
\email{tommaso.mengoni@unicam.it}
\affiliation{University of Camerino, Via Madonna delle Carceri, Camerino, 62032, Italy.}
\affiliation{Istituto Nazionale di Fisica Nucleare (INFN), Sezione di Perugia, Perugia, 06123, Italy.}
\affiliation{INAF - Osservatorio Astronomico di Brera, Milano, Italy.}

\begin{abstract}

We investigate the generation of quantum entanglement between scalar field modes induced by quantized massless tensor perturbations during inflation. Unlike semiclassical approaches, in which metric perturbations act as classical external sources, we quantize both the tensor perturbations and the scalar field, while treating the de Sitter background classically. We describe the tensor perturbation within linearized gravity theory and show, through a scalar-vector-tensor decomposition, that only the two transverse-traceless graviton polarizations represent physical propagating degrees of freedom. We then consider the process $h\rightarrow\phi\phi$, where the scalar particle production represents the source of quantum correlations in momentum space, and derive the corresponding first-order transition amplitude. This process may provide a second-order quantum correction to the gravitational particle production mechanism. We compute the number density of produced particles by selecting only super-Hubble scales, regularizing the production and improving the predictability of the model for the classical particle background. We then study the particle number spectrum in terms of both initial and final particle states. Differently, the entanglement analysis considers scalar modes lying on opposite sides of the Hubble horizon. We compute the reduced density matrices of the resulting tripartite system and distinguish the entanglement of one scalar sector with the remaining degrees of freedom from the scalar--scalar entanglement associated with the particle production process. We quantify the former through the von Neumann entropy density, and the latter considering the negativity. Our results provide a perturbative framework for studying momentum space quantum correlations generated by quantized gravitational perturbations during inflation.

\end{abstract}

\maketitle
\tableofcontents

\section{Introduction}{\label{sec:intro}}

Quantum particle production in an expanding universe \cite{Birrell:1982ix,Ford:2021syk,RevModPhys.96.045005} is one of the main predictions of quantum field theory in curved spacetime. 
In this framework, quantum fields evolve on a classical geometry, whose time dependence can lead to a nontrivial mixing of positive- and negative-frequency modes and hence to particle creation \cite{PhysRevLett.21.562, Parker:1969au, Parker:1971pt}.

The first studies in this field considered unperturbed expanding spacetimes \cite{PhysRevLett.21.562, Parker:1969au,PhysRevD.17.964}, usually described by a Friedmann-Lemaître-Robertson-Walker (FLRW) metric. In these frameworks, a time-dependent background can yield particle-antiparticle pair production from vacuum, and the final particle abundance can be computed through the Bogoliubov transformations. This mechanism, known as the Parker effect, has been investigated in several cosmological contexts, including dark-matter production \cite{PhysRevD.59.023501,PhysRevD.64.043503, PhysRevD.67.083514, PhysRevD.93.103520, Kannike:2016jfs, PhysRevD.96.103540, Fairbairn:2018bsw, PhysRevD.59.023501, Ema:2018ucl, Belfiglio:2024xqt, Belfiglio:2025chv}, reheating \cite{PhysRevD.64.023509, Feng:2002nb, Chun:2009yu, Bettoni:2021zhq}, baryogenesis \cite{Bambi:2006hp,Hashiba:2019mzm,Bernal:2021kaj,PhysRevD.106.075006,PhysRevD.107.063537, Dubbini:2025hcw, Dubbini:2025jjz, Dubbini:2025xvg} and even primordial black hole generation \cite{Erfani:2015rqv,Hooper:2019gtx}. 

The presence of some inhomogeneities may enhance gravitational particle production (GPP) processes from vacuum \cite{Frieman:1985fr, Cespedes:1989kh, PhysRevD.45.4428, Belfiglio:2024swy}. This mechanism is particularly relevant in the early universe, where the accelerated cosmological expansion occurring during inflation \cite{Ford:1986sy,PhysRevD.37.3428,PhysRevD.59.023501,PhysRevD.101.083516,PhysRevD.101.123522} provides a natural amplification of quantum fluctuations and particle production \cite{Frieman:1985fr, Ford:1986sy, Cespedes:1989kh, Ford:2021syk}. In inflationary settings, metric perturbations associated with quantum inflaton fluctuations can provide relevant contributions to the final abundance of gravitationally produced particles \cite{Belfiglio:2022qai, Belfiglio:2023rxb}.

Further, GPP processes may also generate quantum entanglement in the final state of the fields in question \cite{Ball:2005xa,Fuentes:2010dt, Martin-Martinez:2012chf}. Studying these quantum correlations in momentum space \cite{PhysRevD.86.045014, Brahma:2023hki, Mazur:2008wa} can furnish information about the universe dynamics, quantum field statistics, and the quantum-to-classical transition of primordial perturbations \cite{Kiefer:1998qe, Campo:2005sv,Nambu:2008my,Burgess:2006jn,PhysRevD.102.023512,PhysRevD.102.043529,Martin:2015qta, Nelson:2016kjm, Brahma:2021mng, Gong:2019yyz}. Even though direct measurements of entanglement are currently unfeasible, a protocol known as entanglement harvesting \cite{Reznik:2002fz,PhysRevD.92.064042,Barcelo:2005fc,Wilson:2011rsw, Hung:2012nc,PhysRevLett.123.180502} has been developed to extract entanglement from quantum fields exploiting local couplings with simple quantum systems, e.g., Unruh-DeWitt detectors \cite{PhysRevD.14.870,PhysRevD.29.1047,Vezzoli2019, Steinhauer:2021fhb}.

In the present work, we investigate a different mechanism for generating quantum correlations during inflation. We keep the homogeneous de Sitter geometry classical, while quantizing its gravitational spin-2 perturbation within linearized gravity\footnote{Other applications of quantum-field approaches to primordial gravitational perturbations can be found, for instance, in Refs. \cite{Luongo:2025ovo, Luongo:2026ndg, Mengoni:2026clb, Collins:2016ahj, Kanno:2022kve, Ning:2023ybc}.}. This provides a consistent low-energy framework for describing quantum gravitational excitations without invoking a complete theory of quantum gravity. We then consider the graviton interaction with a real scalar field\footnote{For more details about gravitational perturbations, see Ref. \cite{Mukhanov:1990me}.}, and investigate the process $h\to\phi\phi$. Unlike semiclassical treatments, where gravitational perturbations act as external classical fields, and particles are \emph{de facto} produced by vacuum, here we consider an initial graviton state. Accordingly, quantized metric perturbations can directly contribute to the particle production process. The resulting particle abundance may be interpreted as a quadratic quantum correction to the overall GPP mechanism, as long as the linearized theory of gravity is quadratic in $h_{\mu\nu}$.

Within semiclassical setups, inhomogeneous scalar metric perturbations are commonly treated as the main sources of translational invariance breaking, allowing particle pair production with nonzero total final momentum. Vector and tensor perturbations are commonly neglected. By contrast, within our fully quantum scenario, we verify through a scalar-vector-tensor (SVT) decomposition that only the two transverse and traceless tensor perturbations are gauge-invariant, and thus embed the physical degrees of freedom. We show that fixing the gauge completely eliminates all the scalar and vector modes.

Accordingly, we only consider the transverse-traceless graviton tensor modes, and derive the corresponding first-order production amplitude for the process $h\to\phi\phi$. We first derive the transition amplitude in a finite comoving volume and subsequently take the continuum limit. We identify a restricted graviton momentum domain in which the integrated transition weight remains small, guaranteeing perturbative consistency. Moreover, in order to avoid ultraviolet (UV) sensitivity, the particle production estimate is limited to a super-Hubble domain, in which the relevant modes remain well outside the Hubble radius. This specific regularization prescription is natural in cosmological frameworks, and should also improve the predictability of the theory with respect to possible measurements of the classical particle background. Firstly, we discuss the dependence of the transition probability on the graviton momentum, in order to characterize the production in terms of the initial particle state, and then we show the momentum spectrum concerning the scalars, in order to study the main features of the produced particles. Finally, according to Planck data \cite{Planck:2018jri}, we plot the comoving number density of particles as a function of the inflationary energy scale.

The particle production amplitude is then used to characterize the quantum correlations of the produced state. We consider a momentum space partition with respect to a reference Hubble scale \cite{Brahma:2023hki,Rai:2020edx, Belfiglio:2022yvs, Belfiglio:2023moe, Belfiglio:2025cst, Belfiglio:2025ofg, Belfiglio:2026umv}, obtaining a tripartite graviton-scalar-scalar system. As in the particle production, the entanglement analysis also focuses on specific momentum regions. In particular, the scalar momenta are chosen on opposite sides of a reference Hubble scale, while the graviton momentum takes values both below and above the horizon. Within this framework, we distinguish the entanglement of a scalar sector with the remaining degrees of freedom from the entanglement internal to the produced scalar pair. About the former, we compute the von Neumann entropy density, showing in particular how the correlations are distributed with respect to the final particle momenta. We also relate the entanglement entropy to the inflationary energy scale, so that the entanglement analysis may contribute to the predictability of our model. Finally, we investigate the internal scalar-scalar entanglement through the negativity density and the logarithmic negativity with respect to a fixed reference scale, showing remarkably that these two are related with each other.

The paper is organized as follows. In Sect. \ref{sec:LinGrav-flat}, we review the linearized gravity theory in Minkowski spacetime. In Sect. \ref{sec:LinGrav-curved}, we generalize it to a conformally flat de Sitter background and derive the corresponding SVT mode equations. Sect. \ref{sec:curved-production} focuses on scalar particle production during inflation from massless spin-2 particles. There, we also discuss the motivations for restricting our analysis to super-Hubble scales. Then, in Sect. \ref{sec:entanglement}, we investigate the entanglement generation within this process, and, finally, in Sect. \ref{sec:conclusions}, we summarize our results and discuss their implications.

\section{The linearized gravity theory in flat spacetime}{\label{sec:LinGrav-flat}}

A free spin-2 particle in flat spacetime is represented by a tensor field $h_{\mu\nu}$. It follows from expanding the Einstein-Hilbert action to quadratic order in the metric perturbation around Minkowski spacetime $\eta_{\mu\nu}=\text{diag}(+1,-1,-1,-1)$, according to $\eta_{\mu\nu}\to \eta_{\mu\nu}+\frac{2}{\overline{M}_{\text{Pl}}}h_{\mu\nu}$, with $\overline{M}_{\text{Pl}}$ the reduced Planck mass. In this way, we obtain the Lagrangian in flat spacetime, that is
\begin{equation}
\begin{split}
\mathcal{L}_{\text{h}}^{(\text{flat})}=&\frac{1}{2}\partial_{\alpha}h^{\mu\nu}\partial^{\alpha}h_{\mu\nu}-\partial_{\alpha}h^{\mu\nu}\partial_{\nu}h^{\alpha}{}_{\mu}\\&+\partial_{\mu}h^{\mu\nu}\partial_{\nu}h-\frac{1}{2}\partial_{\mu}h\partial^{\mu}h,
\end{split}
\label{eq_PF-kin-lagr-flat}
\end{equation}
with $h\equiv \eta^{\mu\nu}h_{\mu\nu}$. Eq. \eqref{eq_PF-kin-lagr-flat} describes a free massless spin-2 particle in flat spacetime. It is possible to show that under the coordinate change $x^{\mu}\to x^{\mu}+\xi^{\mu}(x)$ the tensor field changes according to $h_{\mu\nu}\to h_{\mu\nu}+\partial_{\mu}\xi_{\nu}+\partial_{\nu}\xi_{\mu}$, and Eq. \eqref{eq_PF-kin-lagr-flat} is invariant under such a field transformation. This gauge symmetry reflects the fact that only 2 physical degrees of freedom are associated with a massless spin-2 particle, corresponding to the states of helicity $\lambda=\pm 2$.

Interactions with matter naturally emerge when some matter fields are minimally coupled to gravity. For simplicity, we consider a real scalar field $\phi$ minimally coupled to gravity, and call $\mathcal{L}_{\phi}^{(\text{flat})}$ its Lagrangian density in flat spacetime. When we perturb the metric according to $\eta_{\mu\nu}\to \eta_{\mu\nu}+\frac{2}{\overline{M}_{\text{Pl}}}h_{\mu\nu}$, we obtain two terms:
\begin{itemize}
\item[-] the zeroth-order term is exactly $\mathcal{L}_{\phi}^{(\text{flat})}$, and describes a free real scalar field in flat spacetime, while
\item[–] the first-order term outlines an interaction between the metric perturbation -- the graviton -- and the scalar field, through the interaction Lagrangian
\begin{equation}
\mathcal{L}_{\text{int}}^{(\text{flat})}=-\frac{1}{\overline{M}_{\text{Pl}}}T_{\mu\nu}^{(0)}h^{\mu\nu},
\label{eq_INT-lagr-flat}
\end{equation}
where $T_{\mu\nu}^{(0)}$ is the energy-momentum tensor of the free scalar field written with respect to $\eta_{\mu\nu}$.
\end{itemize}
In this work, second-order interactions will be neglected with respect to the zero- and first-order contributions.

Thus, the full Lagrangian reads 
\begin{equation}
\mathcal{L}_{(\text{flat})}=\mathcal{L}_{\text{h}}^{(\text{flat})}+\mathcal{L}_{\phi}^{(\text{flat})}+\mathcal{L}_{\text{int}}^{(\text{flat})}.
\label{eq_full-lagr-flat}
\end{equation}

\subsection{Field quantization in flat spacetime}{\label{subsec:quantization-flat-spacetime}}

Field quantization is now straightforward. Indeed, we recall that the full Lagrangian in Eq. \eqref{eq_full-lagr-flat} is obtained from a perturbative expansion around $\eta_{\mu\nu}$ by a quantity proportional to $\frac{h_{\mu\nu}}{\overline{M}_{\text{Pl}}}$. To ensure the validity of the perturbative expansion, we need to assume that $\left|\frac{h}{\overline{M}_{\text{Pl}}}\right|\ll 1$, and this enables us to consider the interaction Lagrangian as a perturbation. Therefore, we can consider both the graviton and the scalar field as free fields, and for them canonical quantization can be used. Any amplitude one may want to compute will be found within the interaction picture, considering the free quantum fields and the $\mathcal{S}-$matrix expanded to first-order in the interaction Lagrangian.

We can write the general mode solution as $h_{\mu\nu}\sim\varepsilon^{(\lambda)}_{\mu\nu}({\bf k})e^{-ik\cdot x}$, where $\varepsilon_{\mu\nu}^{(\lambda)}$ is the polarization tensor. Promoting the field to an operator, such that $h_{\mu\nu}\rightarrow\hat h_{\mu\nu}$, the quantized tensor field reads
\begin{equation}
\hat h_{\mu\nu}(t,\mathbf{x})=\sum_{\lambda=\pm2} \int \frac{d^3k}{(2\pi)^{3/2}}\left[\hat{a}_\lambda(\mathbf{k})\varepsilon_{\mu\nu}^{(\lambda)}(\mathbf{k})\frac{e^{-ik\cdot x}}{\sqrt{2\omega_{\mathbf{k}}}} + \text{h.c.}\right],
\label{eq_h-quantum-flat}
\end{equation}
where $\hat{a}_\lambda$ (and its Hermitian conjugate) are the annihilation (and creation) operators, satisfying the canonical commutation relations $[\hat{a}_\lambda(\mathbf{k}),\hat{a}_{\lambda'}^\dagger(\mathbf{k}')]
=
\delta_{\lambda\lambda'}\delta^{(3)}(\mathbf{k}-\mathbf{k}')$.

On the other hand, the scalar field is a simple free real scalar field in flat spacetime, whose quantum expression is thus
\begin{equation}
\hat{\phi}(t,{\bf x})=\int\frac{d^3k}{(2\pi)^{3/2}}\left[\hat{b}({\bf k})\frac{e^{-ik\cdot x}}{\sqrt{2\omega_{\bf k}}}+\text{h.c.}\right],
\label{eq_phi-quantum-flat}
\end{equation}
with $\left[\hat{b}({\bf k}),\hat{b}^{\dagger}({\bf k'})\right]=\delta^{(3)}({\bf k}-{\bf k'})$.

\section{The linearized gravity theory in curved spacetime}{\label{sec:LinGrav-curved}}

The objective of this work is to study the production of scalar particles and the corresponding entanglement generation from massless gravitons during the inflationary stage. For this purpose, the flat scenario analyzed in Sect. \ref{sec:LinGrav-flat} does not represent a realistic background, as we know that the universe undergoes an accelerated expansion during inflation. Hence, generalizing the treatment to curved spacetime is fundamental \cite{Higuchi:1991tn, Deser:2006sq}.

We consider a conformally flat universe, with metric tensor $g_{\mu\nu}=a^2(\eta)\eta_{\mu\nu}$, with $\eta$ the conformal time, in lieu of the Minkowski metric $\eta_{\mu\nu}$. In addition, we assume that the inflationary period is described by a de Sitter phase, along which the scale factor can be written in terms of the conformal time $\eta$ as $a(\eta)=-\frac{1}{H_I\eta}$, where $H_I$ is the Hubble parameter during inflation. Specifically, we take the benchmark value $\frac{H_I}{\overline{M}_{\text{Pl}}}\simeq 2.50\times 10^{-5}$, in agreement with current Planck constraints \cite{Planck:2018jri}.

We consider the inflationary period going from $\eta_{\text{i}}$ to $\eta_{\text{f}}$. Their values can be found from the pivot scale $k_\text{piv}\simeq 3.15\times 10^{-40}\text{ GeV}$, the number of e-folds after which the pivot scale exits the horizon, $N_\text{piv}=10$, and the total number of e-folds, $N_{\text{f}}=60$. These parameters yield $\eta_{\text{i}}\simeq -6.99\times 10^{43}\text{ GeV}^{-1}$ and $\eta_{\text{f}}\simeq -6.12\times 10^{17}\text{ GeV}^{-1}$. Hence, the crossing-horizon scales respectively at the beginning and the end of inflation are $k_{\text{i}}=-\frac{1}{\eta_{\text{i}}}\simeq 1.43\times 10^{-44}\text{ GeV}$ and $k_{\text{f}}=-\frac{1}{\eta_{\text{f}}}\simeq 1.63\times 10^{-18}\text{ GeV}$.

Generalizing the Lagrangian density in Eq. \eqref{eq_PF-kin-lagr-flat} to a conformally flat de Sitter universe is not a trivial operation. Indeed, in order to describe the propagation of a massless spin-2 particle on a curved background, the most natural approach would seemingly be to take Eq. \eqref{eq_PF-kin-lagr-flat} and promote the ordinary derivatives to covariant derivatives, $\partial_{\mu}\to \nabla_{\mu}$, together with replacing the background metric from flat to curved, $\eta_{\mu\nu}\to g_{\mu\nu}=a^2\eta_{\mu\nu}$. This approach has been proved to be inconsistent in curved spacetime.
\begin{itemize}
\item[-] In the language of symmetries, the generalization of the massless linearized gravity Lagrangian would not be invariant under the field transformation $h_{\mu\nu}\to h_{\mu\nu}+\nabla_{\mu}\xi_{\nu}+\nabla_{\nu}\xi_{\mu}$, corresponding to the coordinate transformation $x^{\mu}\to x^{\mu}+\xi^{\mu}(x)$.
\item[-] This means that a massless graviton in curved spacetime would have more degrees of freedom than a massless graviton in flat spacetime, which is impossible. The gauge invariance of the Lagrangian density is thus a fundamental requirement for the consistency of the theory, and a correct generalization to curved spacetime should preserve it.
\end{itemize}
The theory can be consistently formulated only in Einstein spacetime, for which $R_{\mu\nu}=\lambda g_{\mu\nu}\implies R=4\lambda$, with $\lambda$ a constant. This is actually the case of the de Sitter inflationary universe, where $R_{\mu\nu}=-3H_I^2g_{\mu\nu}$. The derivative part of the curved-background quadratic action can be regarded as the covariantized counterpart of the flat-space Lagrangian in Eq. \eqref{eq_PF-kin-lagr-flat}, obtained through $\eta_{\mu\nu}\rightarrow g_{\mu\nu}$ and $\partial_\mu\rightarrow\nabla_\mu$, yielding
\begin{equation}
\begin{aligned}
\mathcal{L}_{h}^{(\mathrm{flat})}
\rightarrow
\mathcal{L}_{h}^{(\mathrm{cov})}
={}&
\sqrt{-g}\Bigg[
\frac{1}{2}
\nabla_{\alpha}h^{\mu\nu}\nabla^{\alpha}h_{\mu\nu}
-\nabla_{\alpha}h^{\mu\nu}\nabla_{\nu}h^{\alpha}{}_{\mu}
\\
&\qquad\,\,\,\,\,\,
+\nabla_{\mu}h^{\mu\nu}\nabla_{\nu}h
-\frac{1}{2}\nabla_{\mu}h\nabla^{\mu}h
\Bigg].
\end{aligned}
\label{...}
\end{equation}
However, this does not reproduce the complete quadratic Einstein-Hilbert action. Indeed, covariant derivatives do not commute on a curved background, and curvature-dependent terms are required to preserve the linearized gauge symmetry.
To restore the gauge invariance of the massless graviton Lagrangian in curved spacetime, an extra term proportional to the scalar curvature, naturally arising from the quadratic expansion of the Einstein-Hilbert action, appears \cite{Bastianelli:2022pqq}.
Specifically, for the Einstein background considered here, the quadratic massless spin-2 Lagrangian can be written as
\begin{equation}
\mathcal{L}_{\text{h}}^{(\text{flat})}\to \mathcal{L}_{\text{h}}^{(\text{curved})}=\mathcal{L}_{\text{h}}^{(\text{cov})}-\sqrt{-g}\left(h^{\mu\nu}h_{\mu\nu}-\frac{h^2}{2}\right)\frac{R}{4}.
\label{eq_PF-lagr-curved}
\end{equation}
We emphasize that, in an Einstein spacetime, the Lagrangian density in Eq. \eqref{eq_PF-lagr-curved} guarantees the correct number of degrees of freedom, preserving gauge symmetry \cite{Deser:2006sq}. Specifically, for massless spin-2 fields, a convenient gauge choice is the transverse-traceless (TT) gauge, that allows us to satisfy the transversality and tracelessness conditions. We will follow this approach throughout our work.

As suggested in Sect. \ref{sec:LinGrav-flat}, we solve both the scalar and tensor field dynamics neglecting their interactions, and then we compute the amplitude at first order in perturbation theory within the interaction picture. In particular, using the transversality and tracelessness conditions -- satisfied in TT gauge -- the equations of motion of the tensor field get simplified as follows:
\begin{equation}
\begin{split}
&\nabla^2h_{\alpha\beta}-2R_{\alpha\lambda\mu\beta}h^{\lambda\mu}=0\implies \nabla^2h_{\alpha\beta}+2H_I^2 h_{\alpha\beta}=0.
\end{split}
\label{eq_EOM-h-curved}
\end{equation}

Expanding the Laplacian operator in a de Sitter background, we compute the equations of motion for $h_{00}$, $h_{0i}$ and $h_{ij}$, obtaining, respectively
\begin{subequations}
\begin{align}
    &\Box h_{00}+2\mathcal{H}h'_{00}+\left[2a^2H_I^2-2\mathcal{H}^2\right]h_{00}=0,
\label{eq_EOM-h00-massive}\\
&\Box h_{0i}+\left[2a^2H_I^2-4\mathcal{H}^2\right]h_{0i}=-2\mathcal{H}\partial_ih_{00},
\label{eq_EOM-h0i-massive}\\
&\Box h_{ij}-2\mathcal{H}h'_{ij}+\left[2a^2H_I^2-4\mathcal{H}^2\right]h_{ij}\notag\\&\quad=2\mathcal{H}^2\delta_{ij}h_{00}-2\mathcal{H}\left(\partial_ih_{0j}+\partial_jh_{0i}\right),
\label{eq_EOM-hij-massive}
\end{align}
\end{subequations}
where $\mathcal{H}=-\frac{1}{\eta}=aH_I$ is the conformal Hubble parameter. In addition, for the sake of completeness, we write the generalization to curved spacetime of the transversality condition,
\begin{equation}
\nabla^{\mu}h_{\mu\nu}=0\implies \begin{cases}
h'_{00}+2\mathcal{H}h_{00}=\partial_jh_{0j} & \nu = 0\\ h'_{0i}+2\mathcal{H}h_{0i}=\partial_jh_{ij} & \nu = i
\end{cases},
\label{eq_transv-explicit}
\end{equation}
and the tracelessness condition,
\begin{equation}
h=0\implies h_{00}=\delta^{ij}h_{ij}.
\label{eq_tracelessness-explicit}
\end{equation}
As anticipated above and explicitly shown below, the massless gauge symmetry allows us to work in the transverse-traceless gauge. Hence, the tensor equations of motion reduce to Eq. \eqref{eq_EOM-h-curved}.

\subsection{Scalar-vector-tensor decomposition}{\label{sec:SVT}}

Now, we can solve the equations of motion for the tensor field, Eqs. \eqref{eq_EOM-h00-massive}--\eqref{eq_EOM-hij-massive}, respectively for the $00$-, $0i$- and $ij$-components. We solve the equations using the SVT decomposition for $h_{\mu\nu}$, through which a tensor field is separated into irreducible scalar, transverse vector, and transverse-traceless tensor components \cite{Mukhanov:1990me}. In particular, in Fourier space we have
\begin{subequations}
\begin{align}
    &h_{00}(\eta,{\bf x})\sim A_{\bf k}(\eta)e^{i{\bf k}\cdot{\bf x}},
\label{eq_h00-SVT}\\
&h_{0i}(\eta,{\bf x})\sim \left[ik_iB_{\bf k}(\eta)+\sum_{\lambda=\pm 1}\epsilon_i^{(\lambda)}({\bf k})V_{\bf k}^{(\lambda)}(\eta)\right]e^{i{\bf k}\cdot{\bf x}},
\label{eq_h0i-SVT}\\
&h_{ij}(\eta,\mathbf{x})\sim\Bigg[\delta_{ij}\psi_{\bf k}(\eta)-k_i k_j E_{\bf k}(\eta)+\sum_{\lambda=\pm2}
\epsilon_{ij}^{(\lambda)}(\mathbf{k})h_{\bf k}^{(\lambda)}(\eta)\notag\\&+ik_i\sum_{\lambda=\pm1}\epsilon_j^{(\lambda)}(\mathbf{k})F_{\bf k}^{(\lambda)}(\eta)
+ik_j\sum_{\lambda=\pm1}\epsilon_i^{(\lambda)}(\mathbf{k})F_{\bf k}^{(\lambda)}(\eta)\Bigg]e^{i{\bf k}\cdot{\bf x}}.
\label{eq_hij-SVT}
\end{align}
\end{subequations}
The polarization vectors $\epsilon_{i}^{(\lambda)}({\bf k})$ enjoy the following properties:
\begin{subequations}
\begin{align}
&k_i\epsilon_i^{(\lambda)}({\bf k})=0&&\text{(transversality)},
\label{eq_propr-transv-pol-vect}\\
&\epsilon_i^{(\lambda)}({\bf k})\epsilon_i^{(\lambda')*}({\bf k})=\delta_{\lambda\lambda'}&&\text{(orthonormality)},
\label{eq_propr-ortho-pol-vect}\\
&\sum_{\lambda=\pm 1}\epsilon_i^{(\lambda)}({\bf k})\epsilon_j^{(\lambda)*}({\bf k})=\Pi_{ij}({\bf k})&&\text{(completeness)},
\label{eq_propr-compl-pol-vect}
\end{align}
\end{subequations}
with $\Pi_{ij}({\bf k})\equiv \delta_{ij}-\frac{k_ik_j}{|{\bf k}|^2}$.

The polarization tensors $\epsilon_{ij}^{(\lambda)}({\bf k})$ are symmetric and enjoy the following properties:
\begin{subequations}
\begin{align}
&k_i\epsilon_{ij}^{(\lambda)}({\bf k})=k_j\epsilon_{ij}^{(\lambda)}({\bf k})=0&&\text{(transversality)},
\label{eq_propr-transv-pol-tens}\\
&\delta_{ij}\epsilon_{ij}^{(\lambda)}({\bf k})=0&&\text{(tracelessness)},
\label{eq_propr-trace-pol-tens}\\
&\epsilon_{ij}^{(\lambda)}({\bf k})\epsilon_{ij}^{(\lambda')*}({\bf k})=\delta_{\lambda\lambda'}&&\text{(orthonormality)},
\label{eq_propr-ortho-pol-tens}\\
&\sum_{\lambda=\pm 2}\epsilon_{ij}^{(\lambda)}({\bf k})\epsilon_{lm}^{(\lambda)*}({\bf k})=P_{ij,lm}({\bf k})&&\text{(completeness)},
\label{eq_propr-compl-pol-tens}
\end{align}
\end{subequations}
with the projector defined as
\begin{equation}
\begin{split}
P_{ij,lm}({\bf k})\!\equiv \frac{\!\Pi_{il}({\bf k})\Pi_{jm}({\bf k})\!+\!\Pi_{im}({\bf k})\Pi_{jl}({\bf k})\!-\!\Pi_{ij}({\bf k})\Pi_{lm}({\bf k})}{2}.
\end{split}
\label{eq_projector-massive}
\end{equation}
In this way, we have $A_{\bf k}$, $B_{\bf k}$, $\psi_{\bf k}$ and $E_{\bf k}$ as scalar modes, $V_{\bf k}$ and $F_{\bf k}$ as transverse vector modes and $h_{\bf k}$ as the transverse-traceless tensor mode. Eqs. \eqref{eq_EOM-h00-massive}--\eqref{eq_EOM-hij-massive} will give rise to equations for these modes.

\subsection{Gauge fixing}{\label{subsec:gauge}}

When the graviton mass is exactly zero, the theory is invariant under the coordinate transformation $x^\mu\to x^\mu+\xi^\mu(x)$, encoding the field transformation $h_{\mu\nu}\to h_{\mu\nu}+\nabla_\mu\xi_\nu+\nabla_\nu\xi_\mu$. This invariance reflects a gauge symmetry that reduces the physical degrees of freedom to 2. The latter are embedded in the two transverse-traceless tensor modes, labeled with $h_{\bf k}^{(\lambda)}$, for $\lambda=\pm 2$. Indeed, under the field transformation $h_{\mu\nu}\to h_{\mu\nu}+\delta h_{\mu\nu}$, with
\begin{subequations}
\begin{align}
&\delta h_{00}=2\left(\xi_0'-\mathcal{H}\xi_0\right),
\label{eq_transform-h00}\\
&\delta h_{0i}=\partial_0\xi_i+\partial_i\xi_0-2\mathcal{H}\xi_i,
\label{eq_transform-h0i}\\
&\delta h_{ij}=\partial_i\xi_j+\partial_j\xi_i-2\delta_{ij}\mathcal{H}\xi_0,
\label{eq_transform-hij}
\end{align}
\end{subequations}
it is not difficult to prove that the only gauge-invariant modes are $h_{\bf k}^{(\lambda)}$. This can be done by considering the SVT decomposition for $\xi^\mu$, that is,
\begin{subequations}
\begin{align}
&\xi_0(\eta,{\bf x})\sim T_{\bf k}(\eta)e^{i{\bf k}\cdot{\bf x}},
\label{eq_csi0-modes}\\
&\xi_i(\eta,{\bf x})\sim \left[ik_iL_{\bf k}(\eta)+\sum_{\lambda=\pm1}\epsilon_i^{(\lambda)}(k)\xi_{\bf k}^{(\lambda)}(\eta)\right]e^{i{\bf k}\cdot{\bf x}},
\label{eq_csii-modes}
\end{align}
\end{subequations}
yielding straightforwardly $\delta h_{\bf k}^{(\lambda)}=0$.

Conversely, scalar and vector modes transform non-trivially under the gauge transformation, and thus they are purely gauge modes that do not carry any physical information. The corresponding $8$ degrees of freedom should be eliminated by
\begin{itemize}
\item[-] fixing the gauge, and then
\item[-] using the remaining gauge freedom in order to impose that these modes do not propagate
\end{itemize}
Conveniently, we work in the so called transverse and traceless (TT) gauge, that is,
\begin{equation}
\begin{cases}
\nabla^\mu h_{\mu\nu}=0\\ h=0
\end{cases}
\label{eq_gauge-choice},
\end{equation}
\emph{de facto} reproducing the transversality and tracelessness conditions in Eqs. (\ref{eq_transv-explicit}) and (\ref{eq_tracelessness-explicit}). Differently from the massive case, when these two arise directly from the equations of motion, here they can be recovered as a gauge choice. These conditions are necessary in order the equations of motion to get simplified to Eqs. (\ref{eq_EOM-h00-massive})--(\ref{eq_EOM-hij-massive}).

In terms of the Fourier modes, the transversality condition reads
\begin{subequations}
\begin{align}
&A_k'+2\mathcal{H}A_k=-k^2B_k
\label{eq_0-transv-Fourier} && \nu=0,\\
&B_k'+2\mathcal{H}B_k=\psi_k-k^2E_k&&\nu=i\quad (\text{scalar sector}),
\label{eq_i-transv-scalar}\\
&V_k^{(\lambda)}{'}+2\mathcal{H}V_k^{(\lambda)}=-k^2F_k^{(\lambda)}&&\nu=i\quad (\text{vector sector}),
\label{eq_i-transv-vector}
\end{align}
\end{subequations}
while the tracelessness condition becomes
\begin{equation}
A_{\bf k}=3\psi_{\bf k}-k^2E_{\bf k}.
\label{eq_tracelessness-modes}
\end{equation}
Overall, fixing the gauge removes $5$ degrees of freedom, leaving still $3$ unphysical ones. In this respect, we use the remaining gauge freedom in order to set $A_{\bf k}=0$ and $V_{\bf k}^{(\lambda)}=0$, for $\lambda=\pm 1$. Accordingly, the transversality and tracelessness conditions imply that all the scalar and vector modes vanish identically, confirming that for a massless graviton none of them can propagate.

In view of that, we are interested only in the dynamics of $h_{\bf k}^{(\lambda)}$. It can be derived from the tensor sector of Eq. (\ref{eq_EOM-hij-massive}), namely
\begin{equation}
h_{{\bf k}}^{(\lambda)}{''}\!\!-2\mathcal{H}h_{\bf k}^{(\lambda)}{'}\!+\!\left[k^2-\frac{\nu^2-\frac{1}{4}}{\eta^2}\right]h_{\bf k}^{(\lambda)}=0,\quad \quad \nu=\frac{3}{2}.
\label{eq_tensor-sector-hij}
\end{equation}
This can be solved in a simple way with respect to $\tilde{h}_{\bf k}^{(\lambda)}\equiv \frac{h_{\bf k}^{(\lambda)}}{a}$, yielding
\begin{equation}
h_{\bf k}^{(\lambda)}(\eta)=a(\eta)\frac{\sqrt\pi}{2}e^{i\frac{\pi}{2}\left(\nu+\frac{1}{2}\right)}\sqrt{-\eta}H^{(1)}_{\nu}(-k\eta),
\label{eq_h-sol-massive}
\end{equation}
that is, specialized for $\nu=\frac32$,
\begin{equation}
h_{\bf k}^{(\lambda)}(\eta)=\frac{e^{-ik\eta}}{\sqrt{2k}}\left(1-\frac{i}{k\eta}\right)a(\eta).
\label{eq_h-sol-massless}
\end{equation}
The unique dynamical physical quantities are the tensor modes $h_{\bf k}^{(\lambda)}$, which are gauge-invariant and carry the only two physical degrees of freedom of the theory. Only for completeness, we report in Appendix \ref{app_modedyamics} the equations governing the dynamics of scalar and vector modes.

The quantization of the tensor field thus proceeds by accounting only for the two physical degrees of freedom coming from the transverse-traceless tensor modes \cite{Allen:1986tt}. In particular, we can write
\begin{equation}
\hat{h}_{ij}(\eta,{\bf x})\!=\!\!\!\sum_{\lambda=\pm 2}\int\!\!\frac{d^3k}{(2\pi)^{\frac{3}{2}}}\left[\hat{a}_{\lambda}({\bf k})\epsilon_{ij}^{(\lambda)}({\bf k})h_{\bf k}^{(\lambda)}(\eta)e^{i{\bf k}\cdot{\bf x}}+\text{h.c.}\right],
\label{eq_quantum-tensor-field-massless}
\end{equation}
with $\left[\hat{a}_{\lambda}({\bf k}),\hat{a}^{\dag}_{\lambda'}({\bf p})\right]=\delta^{(3)}\left({\bf k}-{\bf p}\right)\delta_{\lambda\lambda'}$, and the polarization tensors satisfying all the properties in Eqs. \eqref{eq_propr-transv-pol-tens}--\eqref{eq_propr-compl-pol-tens}.

\subsection{Generalization to curved spacetime of the interacting Lagrangian}{\label{subsec:matter-curved}}

Once understood how to generalize the massless spin-2 Lagrangian to curved spacetime, we can do the same for the whole Eq. \eqref{eq_full-lagr-flat}. In this respect, we can simply write
\begin{subequations}
\begin{align}
&\mathcal{L}_{\phi}^{(\text{flat})}\to \mathcal{L}_{\phi}^{(\text{curved})}=\sqrt{-g}\left[\frac{1}{2}\nabla^{\mu}\phi\nabla_{\mu}\phi-\frac{1}{2}m^2_{\phi}\phi^2\right],
\label{eq_phi-lagr-curved}\\
%\end{equation}
%\begin{equation}
&\mathcal{L}_{\text{int}}^{(\text{flat})}\to \mathcal{L}_{\text{int}}^{(\text{curved})}=\sqrt{-g}\left[-\frac{1}{\overline{M}_{\text{Pl}}}T_{\mu\nu}^{(0)}h^{\mu\nu}\right],
\label{eq_INT-lagr-curved}
\end{align}
\end{subequations}
with $T_{\mu\nu}^{(0)}$ now computed with respect to the background metric $g_{\mu\nu}=a^2\eta_{\mu\nu}$. Finally, we obtain
\begin{equation}
\mathcal{L}_{(\text{curved})}=\mathcal{L}_{\text{h}}^{(\text{curved})}+\mathcal{L}_{\phi}^{(\text{curved})}+\mathcal{L}_{\text{int}}^{(\text{curved})}.
\label{eq_full-lagr-curved}
\end{equation}
The quantum scalar field reads
\begin{equation}
\hat{\phi}(\eta,{\bf x})=\int\frac{d^3k}{(2\pi)^{\frac{3}{2}}}\left[\hat{b}({\bf k})\phi_{\bf k}(\eta)e^{i{\bf k}\cdot{\bf x}}+\text{h.c.}\right],
\label{eq_quantum-scalar field}
\end{equation}
with $\left[\hat{b}({\bf k}),\hat{b}^{\dag}({\bf p})\right]=\delta^{(3)}\left({\bf k}-{\bf p}\right)$.

The modes $\phi_{\bf k}$ satisfy the equation
\begin{equation}
\phi_{\bf k}''+2\mathcal{H}\phi_{\bf k}'+\left(k^2+a^2m^2_{\phi}\right)\phi_{\bf k}=0.
\label{eq_EOM-scalar field}
\end{equation}
It is convenient to rewrite Eq. \eqref{eq_EOM-scalar field} in terms of $\tilde{\phi}_{\bf k}\equiv a\phi_{\bf k}$, and setting the usual Bunch-Davies initial conditions we obtain
\begin{equation}
\phi_{\bf k}(\eta)=\frac{1}{a(\eta)}\frac{\sqrt\pi}{2}e^{i\frac{\pi}{2}\left(\nu_{\phi}+\frac{1}{2}\right)}\sqrt{-\eta}H^{(1)}_{\nu_{\phi}}(-k\eta),
\label{eq_scalar field-sol}
\end{equation}
with $\nu_{\phi}\equiv \sqrt{\frac{9}{4}-\frac{m^2_{\phi}}{H_I^2}}$.
For simplicity, we consider massless scalar particles ($\nu_\phi=\frac32$), for which the mode solutions get simplified to
\begin{equation}
\phi_{\bf k}(\eta)=\frac{H_I}{\sqrt{2k^3}}(i-k\eta)e^{-ik\eta}.
\label{eq_phi-sol-massless}
\end{equation}

\section{Graviton-induced scalar particle production}{\label{sec:curved-production}}

Now, we compute the particle production amplitude associated with the process $h\to\phi\phi$. For this purpose, we describe the transition over the finite interval $[\eta_\text{i},\,\eta_\text{f}]$ through the first-order interaction-picture evolution operator. For simplicity, in analogy to the flat scenario, we denote its first-order expansion as the $\mathcal{S}-$matrix coming from the interaction Lagrangian in Eq. \eqref{eq_INT-lagr-curved}. In addition, the quantum fields involved are given by Eq. \eqref{eq_quantum-scalar field} -- for the scalar field -- and Eq. \eqref{eq_quantum-tensor-field-massless} for the unique non-null components of the tensor field. Specifically, we find
\begin{equation}
\begin{split}
\langle\phi({\bf q_1})\phi({\bf q_2})&|\hat{\mathcal{S}}|h({\bf p},\lambda)\rangle=\frac{2i}{\overline{M}_{\text{Pl}}}\frac{(q_1)_i(q_2)_j\epsilon_{ij}^{(\lambda)}(p)}{(2\pi)^{\frac{3}{2}}}\\&\times \delta^{(3)}\left({\bf p}-{\bf q_1}-{\bf q_2}\right)\mathcal{I}_0({\bf q_1},{\bf q_2},{\bf p}),
\end{split}
\label{eq_amplitude-curved-massless}
\end{equation}
with\footnote{The time integration interval will be specified in the following.}
\begin{equation}
\mathcal{I}_0({\bf q_1},{\bf q_2},{\bf p})\equiv\int\phi^*_{\bf q_1}(\eta)\phi^*_{\bf q_2}(\eta)h_{\bf p}^{(\lambda)}(\eta)d\eta.
\label{eq_Imassless}
\end{equation}
The mode solutions are Eq. \eqref{eq_scalar field-sol} for $\phi_{\bf k}$ and Eq. \eqref{eq_h-sol-massless} for $h^{(\lambda)}_{\bf k}$. In particular, we find\footnote{The full calculations can be found in Appendix \ref{app_projectors-momenta}.}
\begin{equation}
\begin{split}
\sum_{\lambda=\pm 2}\!\!\big|\!\langle\hat{\phi}({\bf q_1})&\hat{\phi}({\bf q_2})|\hat{S}|\hat{h}({\bf p},\!\lambda)\rangle\!\big|^2\!\!\!=\!\!\frac{2}{\overline{M}^2_{\text{Pl}}}\frac{V}{(2\pi)^6}\delta^{(3)}\!\!\left({\bf p}\!-\!{\bf q_1}\!-\!{\bf q_2}\right)\! \\&\times \frac{q_1^4q_2^4(1-x^2)^2}{p^4}\left|\mathcal{I}_0({\bf q_1},{\bf q_2},{\bf p})\right|^2.
\end{split}
\label{eq_summed-amplitude2-curved-massless}
\end{equation}
In the semiclassical framework, where the gravitational perturbation is classical, gravitational particle production is addressed by starting from a vacuum state. There, the main production mechanism is based on the expansion of the universe, yielding non-null Bogoliubov coefficients. In that case, an interaction between the gravitational perturbation and the produced field affects the production only as a second-order contribution, while the dominant terms are related to the Bogoliubov coefficients. Conversely, in our framework, since both the gravitational perturbation and the scalar field are quantized, it is natural to understand the particle production in terms of the process $h\to \phi\phi$. This analysis may be understood as a quadratic quantum correction to the standard Parker effect, where quantized excitations of the classical geometry enhance the overall process.

%Unlike semiclassical gravitational particle production, where metric perturbations act as classical sources, our framework treats the graviton as an initial quantum excitation. The process $h\to\phi\phi$ thus provides an additional perturbative channel for scalar particle production. Its contribution is evaluated through the corresponding transition amplitudes as follows.}}

In our fully quantum scenario, the total Hilbert space factorizes as
\begin{equation}
{\cal H}={\cal H}_{h}\otimes{\cal H}_{\phi}\otimes{\cal H}_{\phi},
\label{eq_Hilbert-factorization-production}
\end{equation}
and is thus tripartite. This is a substantial difference with the semiclassical framework, where the system is bipartite.

For a graviton of momentum ${\bf p}$ and helicity $\lambda$, we define the finite volume transition coefficient
\begin{equation}
 C_{{\bf q}_1{\bf q}_2}^{(\lambda)}
 \equiv
 {}_V\langle
 0_h;\phi({\bf q}_1)\phi({\bf q}_2)
 |\hat S|
 h({\bf p},\lambda);0_\phi
 \rangle_V .
\label{eq_C-definition-production}
\end{equation}
There, the subscript $V$ indicates that Eq. (\ref{eq_C-definition-production}) describes the transition coefficient at the discrete level. Indeed, to properly define the continuum limit, we should start by quantizing the fields in a finite comoving volume $V$. Then, the continuum and finite volume normalizations correspond to
\begin{equation}
 \delta^{(3)}({\bf k}-{\bf k}')
 \longrightarrow
 \frac{V}{(2\pi)^3}\delta_{{\bf k},{\bf k}'},
 \qquad
 |{\bf k}\rangle=
 \sqrt{\frac{V}{(2\pi)^3}}\,
 |{\bf k}\rangle_V .
\label{eq_box-normalization-production}
\end{equation}
Using Eq. \eqref{eq_amplitude-curved-massless} and the normalization in Eq. \eqref{eq_box-normalization-production}, we obtain
\begin{equation}
 C_{{\bf q}_1{\bf q}_2}^{(\lambda)}
 =
 \frac{2i}{\overline{M}_{\text{Pl}}\sqrt V}
 (q_1)_i(q_2)_j
 \epsilon_{ij}^{(\lambda)}({\bf p})\,
 {\cal I}_0(q_1,q_2,p)\,
 \delta_{{\bf p},{\bf q}_1+{\bf q}_2}.
\label{eq_C-explicit-production}
\end{equation}
Thus, $C_{{\bf q}_1{\bf q}_2}^{(\lambda)}$ is a dimensionless transition amplitude for a definite momentum channel. In particular, for fixed ${\bf p}$, momentum conservation implies ${\bf q}_2={\bf p}-{\bf q}_1$.

Summing over the two physical graviton helicities, as given in Eq. \eqref{eq_propr-compl-pol-tens}, and using Eq. \eqref{eq_projector+momenta}, we recover the finite volume counterpart of Eq. \eqref{eq_summed-amplitude2-curved-massless}:
\begin{equation}
\begin{split}
 \sum_{\lambda=\pm2}
 \left|C_{{\bf q}_1{\bf q}_2}^{(\lambda)}\right|^2
 =&
 \frac{2}{\overline{M}_{\text{Pl}}^2V}\,
 \delta_{{\bf p},{\bf q}_1+{\bf q}_2}
 \frac{q_1^4q_2^4(1-x^2)^2}{p^4}
 \\
 &\times
 \left|{\cal I}_0(q_1,q_2,p)\right|^2 .
\end{split}
\label{eq_C-squared-production}
\end{equation}
For a fixed incoming graviton mode $({\bf p,\lambda})$, the initial state reads
\begin{equation}
 |\Psi\rangle=
 |h({\bf p},\lambda)\rangle
 |0\rangle_\phi |0\rangle_\phi.
\label{eq_initial-state-production}
\end{equation}
Its first-order evolution can be written as
\begin{equation}
\begin{split}
 |\Psi_{{\bf p},\lambda}\rangle
 =&
 \frac{1}
 {\sqrt{1+{\cal P}_{\bf p}^{(\lambda)}}}
 \Bigg[
 |h({\bf p},\lambda)\rangle
 |0\rangle_\phi |0\rangle_\phi
 \\
 &+
 \sum_{{\bf q_1},{\bf q_2}}
 C_{{\bf q}_1{\bf q}_2}^{(\lambda)}
 |0\rangle_h
 |\phi({\bf q}_1)\rangle
 |\phi({\bf q}_2)\rangle
 \Bigg],
\end{split}
\label{eq_total-state-production}
\end{equation}
where ${\cal P}_{\bf p}^{(\lambda)}$ is given by
\begin{equation}
 {\cal P}_{\bf p}^{(\lambda)}
 \equiv
 \sum_{{\bf q_1},{\bf q_2}}
 \left|C_{{\bf q}_1{\bf q}_2}^{(\lambda)}\right|^2,
\label{eq_pair-probability-production}
\end{equation}
yielding the total particle production probability\footnote{In the perturbative limit with ${\cal P}_{\bf p}\ll1$, we will refer to this quantity as particle production probability.} per incoming graviton mode and polarization ${\cal P}_{\bf p}^{(\lambda)}/(1+{\cal P}_{\bf p}^{(\lambda)})$.

The comoving number density of produced particles turns out to be
\begin{equation}
na^3=\frac{1}{V}\sum_{\lambda=\pm 2}\sum_{{\bf p}}\frac{\mathcal{P}_{\bf p}^{(\lambda)}}{1+\mathcal{P}_{\bf p}^{(\lambda)}}.
\label{eq_n-discrete}
\end{equation}
Remarkably, the relevant quantity is $\mathcal{P}_{\bf p}^{(\lambda)}$. In the continuum limit, it reads
\begin{equation}
\mathcal{P}_{\bf p}^{(\lambda)}=\frac{1}{\overline{M}_\text{Pl}^2}\int \frac{d^3q_1}{(2\pi)^3}\frac{q_1^4q_2^4(1-x^2)^2}{p^4}\left|\mathcal{I}_0(q_1,q_2,p)\right|^2,
\label{eq_Pp-continuum}
\end{equation}
where we assume that $\mathcal{P}_{\bf p}^{(-2)}=\mathcal{P}_{\bf p}^{(+2)}$, and transform the discrete sums into continuous integrals with $\sum_{\bf k}\to \frac{V}{(2\pi)^3}\int d^3k$. In addition, $q_2=|{\bf p}-{\bf q_1}|$ and $x=\frac{{\bf q_1}\cdot{\bf q_2}}{q_1q_2}=\frac{{\bf q_1}\cdot({\bf p}-{\bf q_1})}{q_1|{\bf p}-{\bf q_1}|}=\frac{p^2-q_1^2-q_2^2}{2q_1q_2}$.

For simplicity, the integral $d^3q_1$ may be rewritten as $2\pi q_1^2dq_1d\xi$, with $\xi=\frac{{\bf p}\cdot{\bf q_1}}{pq_1}=\frac{p^2+q_1^2-q_2^2}{2pq_1}$. Then, we notice that $d\xi=-\frac{q_2}{pq_1}dq_2$, and we conveniently change integration variables according to $(q_1,\xi)\to (q_1,q_2)$, with Jacobian determinant $\mathcal{J}=\frac{q_2}{pq_1}$. Finally, we obtain
\begin{equation}
\begin{split}
&\mathcal{P}_{\bf p}^{(\lambda)}(p)=\frac{1}{4\pi^2\overline{M}_\text{Pl}^2}\int_{\mathcal{D}_n }dq_1dq_2\frac{q_1^5q_2^5}{p^5}\left|\mathcal{I}_0(q_1,q_2,p)\right|^2\times\\
&\qquad \qquad \qquad \qquad \times \left[1-x^2(q_1,q_2,p)\right]^2.
\end{split}
\label{eq_Pp-continuum-final}
\end{equation}
The integration domain $\mathcal{D}_n$ deserves a rather thorough analysis. In the following, we will give all the details about the question.

\subsection{Regularization prescription and predictability of the model}{\label{subsec:regularization}}

%In quantum cosmology, particle production is commonly studied by choosing the Bunch-Davies vacuum state, as in our framework. That is, the free field modes are required to approach the Minkowski limit sufficiently far in the past. Even though the Bunch-Davies vacuum state is the standard vacuum choice adopted in the literature, it suffers from strong UV sensitivity in the high-momentum region. Unavoidably, this yields UV divergences when particle production processes are evaluated without setting a proper UV cutoff.

%The same UV behavior also occurs if we compute the number density of produced particles without setting any UV cutoff for the momenta. In fact, the question is not limited to cosmology only, but rather it is a well-known problem in quantum field theory. Therefore, the theory requires an appropriate regularization prescription in order to provide a reliable calculation of the total particle abundance.

In quantum cosmology, particle production is commonly studied by imposing Bunch-Davies initial conditions on the free-field modes, as in our framework. These conditions require the modes to approach their Minkowski counterparts sufficiently far in the past. Although this prescription is standard in the literature, particle-production observables may exhibit strong UV sensitivity when arbitrarily high momenta are included. In particular, computing the number density of produced particles without an appropriate UV cutoff may lead to divergences. This is a familiar issue in quantum field theory, requiring a suitable regularization prescription to obtain a reliable estimate of the particle abundance.

Accordingly, we restrict the analysis to the super-Hubble scales \cite{Kiefer:1998qe, Burgess:2006jn, PhysRevD.102.023512, PhysRevD.102.043529}. This choice is motivated by the following considerations.
\begin{itemize}
\item[-] From the point of view of quantum field theory, this is a choice of the UV cutoff, constraining the momenta to $k\ll-\frac{1}{\eta}$ for each $\eta$ in the time integration interval. We remark that, in this way, we are \emph{de facto} regularizing the particle production process.
%\item[-] In cosmology, super-Hubble modes correspond to the long-wavelength sector, in which primordial perturbations become approximately frozen. Thus, this domain considers the modes that are most directly related to the classical large-scale cosmological perturbations, physically motivating the domain restriction. Therefore, the super-Hubble cutoff represents a natural choice for the UV cutoff in any cosmological frameworks.
%\item[-] In addition, linearized gravity represents the natural low-energy framework for describing quantized tensor perturbations, and is the unique testable theory of quantum gravity. From an observational perspective, we expect that feasible experiments would be sensitive only to the classical particle contribution from the process $h\to\phi\phi$. This classical contribution is only provided by the super-Hubble modes, because the latter are approximately frozen on the Hubble time scale. Therefore, in addition to furnishing a natural regularization prescription for our cosmological problem, restricting the analysis to super-Hubble modes should improve the predictability of the model. At first glance, this consideration suggests that the super-Hubble limit should concern only the produced particles, namely the scalars. Nonetheless, momentum conservation constrains the graviton modes to satisfy the same condition.
\item[-] In cosmology, super-Hubble modes correspond to the long-wavelength sector, where primordial perturbations become approximately frozen. This sector is naturally associated with large-scale cosmological perturbations, providing a physical motivation for the momentum restriction. In addition, linearized gravity provides a well-established low-energy framework for describing quantized tensor perturbations. From an observational perspective, we are particularly interested in the contribution of the process $h\to\phi\phi$ to the large-scale classical particle background. Hence, the super-Hubble cutoff represents a natural regularization choice in the present inflationary framework, allowing us to focus on the contribution most directly related to such a background, thus improving the physical interpretation and potential predictability of the model. Although this consideration concerns the produced scalar particles, momentum conservation also constrains the allowed graviton momenta.
\end{itemize}

For these reasons, in this work we specialize the treatment for super-Hubble scales, and we consider the corresponding approximation for all the modes. For this purpose, we need to guarantee that the condition $-k\eta\ll 1$ is satisfied within all the time and momentum region we integrate over.
\begin{itemize}
\item[-] We select an arbitrary time, say $\eta_*$, and consider only the modes that are super-Hubble within the time interval $[\eta_*,\eta_\text{f}]$.
\item[-] We define the corresponding horizon scale, $k_*\equiv -\frac{1}{\eta_*}$. This is the scale that crosses the horizon at the time $\eta_*$. All the scales lower than $k_*$ have crossed the horizon earlier, and are thus super-Hubble for $\eta>\eta_*$.
\item[-] Consequently, we restrict all the momentum variables within $[k_\text{i},0.01k_*]$.
\end{itemize}
Once we ensure the validity of the super-Hubble condition $-k\eta\ll 1$ within the whole integration domain, we can simplify the modes in this regime. In particular, we have
\begin{subequations}
\begin{align}
&h_{\bf k}^{(\lambda)}(\eta)\xrightarrow{-k\eta\ll 1}\frac{i}{\sqrt{2k^3}}\frac{1}{H_I\eta^2},
\label{eq_h-sol-massless-SH}\\
&\phi_{\bf k}(\eta)\xrightarrow{-k\eta\ll 1}\frac{iH_I}{\sqrt{2k^3}}.
\label{eq_scalar field-mode-SH-massless}
\end{align}
\end{subequations}
Thus, the function $\mathcal{I}_0$ gets simplified to
\begin{equation}
\mathcal{I}_0(q_1,q_2,p)=-\frac{iH_I\left(k_\text{f}-k_*\right)}{(2q_1q_2p)^\frac32}.
\label{eq_I0bar-SH}
\end{equation}

\subsection{Characterization of the momentum integration domain}{\label{subsec:momentum-integration}}

The analysis of the momentum domain $\mathcal{D}_n$ is not yet complete. At this stage, we have $q_1,q_2,p\in[k_\text{i},0.01k_*]$. However, this domain does not guarantee momentum conservation between initial and final states. Indeed, if we assume $q_1\in[k_\text{i},0.01k_*]$, the possible values of $q_2$ should be chosen in a way that both $q_2\in[k_\text{i},0.01k_*]$ and ${\bf q_2}={\bf p}-{\bf q_1}$. In particular, since $\xi \in[-1,1]$ and $q_2=\sqrt{p^2+q_1^2-2pq_1\xi }$, momentum conservation yields $q_2\in[|p-q_1|,p+q_1]$. Overall, we have
\begin{equation}
\mathcal{D}_n:\begin{cases}
k_\text{i}<q_1<0.01k_*\\
\text{max}\left\{k_\text{i},|p-q_1|\right\}<q_2<\text{min}\left\{0.01k_*,p+q_1\right\}
\end{cases}.
\label{eq_integration-domain-production}
\end{equation}
The integration domain $\mathcal{D}_n$ is well-defined if and only if the integration bounds for $q_2$ are well-defined, that is, $\text{max}\left\{k_\text{i},|p-q_1|\right\}<\text{min}\left\{0.01k_*,p+q_1\right\}$. In this regard, sufficient and necessary conditions are $k_\text{i}<0.01k_*$, $k_\text{i}<p+q_1$, $|p-q_1|<p+q_1$ and $|p-q_1|<0.01k_*$. The first three are automatically satisfied, while the fourth one requires $p-0.01k_*<q_1<p+0.01k_*$. However, it is straightforward to prove that the intersection between $[p-0.01k_*,p+0.01k_*]$ and $[k_\text{i},0.01k_*]$ gives $[k_\text{i},0.01k_*]$, considering $p\in[k_\text{i},0.01k_*]$. In this regard, the integration domain $\mathcal{D}_n$ is well-defined.

Then, we need to specify the reference scale $k_*$, which determines both the time interval and the momentum region considered in the calculation. In particular, the higher $k_*$, the smaller the time interval considered, and the wider the super-Hubble momentum region. Accordingly, the choice of the reference scale $k_*$ unavoidably affects the final particle abundance. In this respect, we need to ensure the perturbative regime, that is, to require the transition weight $\mathcal{P}_{\bf p}^{(\lambda)}$ to remain much smaller than unity. In the following, we will show that $\mathcal{P}_{\bf p}^{(\lambda)}$ decreases with the graviton momentum $p$, and thus the process turns out to be perturbative as much as we restrict the analysis to high graviton momenta. We choose $k_*=k_\text{i} e^{N_*}=k_\text{f} e^{N_*-N_\text{f}}$, with $N_*=56$, for which Eq. \eqref{eq_I0bar-SH} simplifies according to $k_\text{f}-k_*\simeq k_\text{f}$. Moreover, we need to further restrict the graviton momentum range to $p\in[0.001k_*,0.01k_*]$, considering only the corresponding portion of the production process. Within this domain, the transition weight remains perturbatively small, consistently with our interpretation of the process as a perturbative quantum correction to gravitational particle production.

\subsection{The number density of particles and momentum spectrum}{\label{subsec:n-spectrum}}

\begin{figure}
\includegraphics[width=0.45\textwidth]{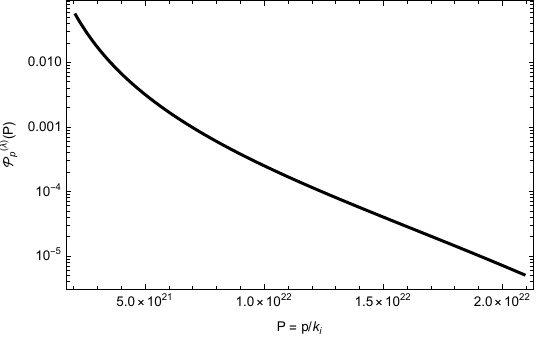}
\caption{The plot is made using a lin-log scale. We show the momentum spectrum per polarization $\mathcal{P}^{(\lambda)}_{\bf p}$ as a function of the rescaled graviton momentum $P=\frac{p}{k_\text{i}}$.}
\label{fig_Pp}
\end{figure}

Firstly, we show that within the momentum domain $\mathcal{D}_n$ and the selected range of graviton momenta, the process really remains perturbative. For this purpose, in Fig. \ref{fig_Pp}
we plot the function $\mathcal{P}_{\bf p}^{(\lambda)}(P)$ with respect to the rescaled momentum $P\equiv\frac{p}{k_\text{i}}$. We can check that it is much smaller than one for all $P$ in the considered range. We remark that this condition is not an arbitrary assumption, but rather it shall be satisfied in order to make the overall treatment consistent. In this regard, Eq. (\ref{eq_n-discrete}) gets simplified to
\begin{equation}
na^3=\frac{1}{\pi^2}\int_{0.001k_*}^{0.01k_*} p^2\mathcal{P}_{\bf p}(p)dp,
\label{eq_n-continuum-final}
\end{equation}
where we consider $\mathcal{P}_{\bf p}^{(-2)}=\mathcal{P}_{\bf p}^{(+2)}\equiv \mathcal{P}_{\bf p}$, and therefore $\mathcal{P}_{\bf p}$ is given by Eq. (\ref{eq_Pp-continuum-final}) independently of the polarization $\lambda$. 

%The function $\mathcal{P}_{\bf p}^{(\lambda)}$ is \emph{de facto} the momentum spectrum concerning the initial graviton state. Therefore, Fig. \ref{fig_Pp} gives us information about the amount of produced particles per graviton momentum and polarization. In particular, we can see that the largest particle production contribution is obtained for low graviton momenta, while a negligible fraction of scalars is produced at the high-momentum end of the spectrum.

The function $\mathcal{P}_{\bf p}^{(\lambda)}$ characterizes the particle production in terms of the initial particle state, that is, in terms of the incoming graviton momentum and polarization. Specifically, as shown in Fig. \ref{fig_Pp}, the transition weight is largest at low graviton momenta and decreases towards the upper end of the selected range. Remarkably, this motivates the choice of $N_*=56$ to preserve the perturbative regime, because for smaller values the considered graviton momentum range would have been shifted to the left, yielding a $\mathcal{P}_{\bf p}^{(\lambda)}$ comparable to unity.

Moreover, we are also interested in investigating the features of the produced particles. For this purpose, we should consider the overall integrand function in Eq. (\ref{eq_n-continuum-final}), and change variables according to $dp=\frac{q_1q_2}{p}dx$. In this way, we are able to completely describe the final particle state in terms of both the scalar particle momenta and their emission angle. We define
\begin{equation}
\begin{split}
&\mathcal{K}_0(q_1,q_2,x)\equiv\frac{1}{k_\text{i}^3}\frac{q_1^6q_2^6\left(1-x^2\right)^2}{p^4(q_1,q_2,x)}\left|\mathcal{I}_0(q_1,q_2,x)\right|^2,
\end{split}
\label{eq_K-function}
\end{equation}
such that
\begin{equation}
na^3=\frac{k_\text{i}^3}{4\pi^4\overline{M}^2_\text{Pl}} \int_{\tilde{\mathcal{D}}_n}dq_1dq_2\int_{x_-(q_1,q_2)}^{x_+(q_1,q_2)} dx\mathcal{K}_0(q_1,q_2,x),
\label{eq_n-K0}
\end{equation}
with
\begin{subequations}
\begin{align}
&x_-(q_1,q_2)\equiv \text{max}\left\{-1,\frac{(0.001k_*)^2-q_1^2-q_2^2}{2q_1q_2}\right\},
\label{eq_xmin}\\
&x_+(q_1,q_2)\equiv\text{min}\left\{1,\frac{(0.01k_*)^2-q_1^2-q_2^2}{2q_1q_2}\right\}
\label{eq_xmax}.
\end{align}
\end{subequations}
The integration bounds $x_-(q_1,q_2)$ and $x_+(q_1,q_2)$ guarantee momentum conservation, further than $x\in[-1,1]$. On the other hand, the integration domain $\tilde{\mathcal{D}}_n$ is seemingly the square $[k_\text{i},0.01k_*]^2$. However, in order $\tilde{\mathcal{D}}_n$ to be well-defined, we should check that $x_-(q_1,q_2)<x_+(q_1,q_2)$ independently of $q_1$ and $q_2$. Proceeding analogously to the case of $\mathcal{D}_n$, the only non-trivial inequality to verify is $q_1+q_2>0.001k_*$. However, this is not automatically satisfied, and therefore we have to restrict the domain $\tilde{\mathcal{D}}$ to
\begin{equation}
\tilde{\mathcal{D}}_n:\begin{cases}
k_\text{i}<q_1<0.01k_*\\ \text{max}\left\{k_\text{i},0.001k_*-q_1\right\}<q_2<0.01k_*
\end{cases}.
\label{eq_domain_Dtilde}
\end{equation}
The function $\mathcal{K}_0(q_1,q_2,x)$ is \emph{de facto} the momentum spectrum concerning the final scalar particle states. In this formulation, we are leaving the scalar momenta $q_1$ and $q_2$ free to vary within the super-Hubble scales, while momentum conservation and the fact that $p\in[0.001k_*,0.01k_*]$ are guaranteed by the integration bounds for $x$ in Eqs. (\ref{eq_xmin}) and (\ref{eq_xmax}). Remarkably, in this case the momentum $p$ is replaced by $p=\sqrt{q_1^2+q_2^2+2q_1q_2x}$, in order to study only the features of the produced scalar particles. Definitely, computing the comoving number density of particles from either Eq. (\ref{eq_n-continuum-final}) or Eq. (\ref{eq_n-K0}) yields the same result.

\begin{figure*}
\includegraphics[width=1.0\textwidth]{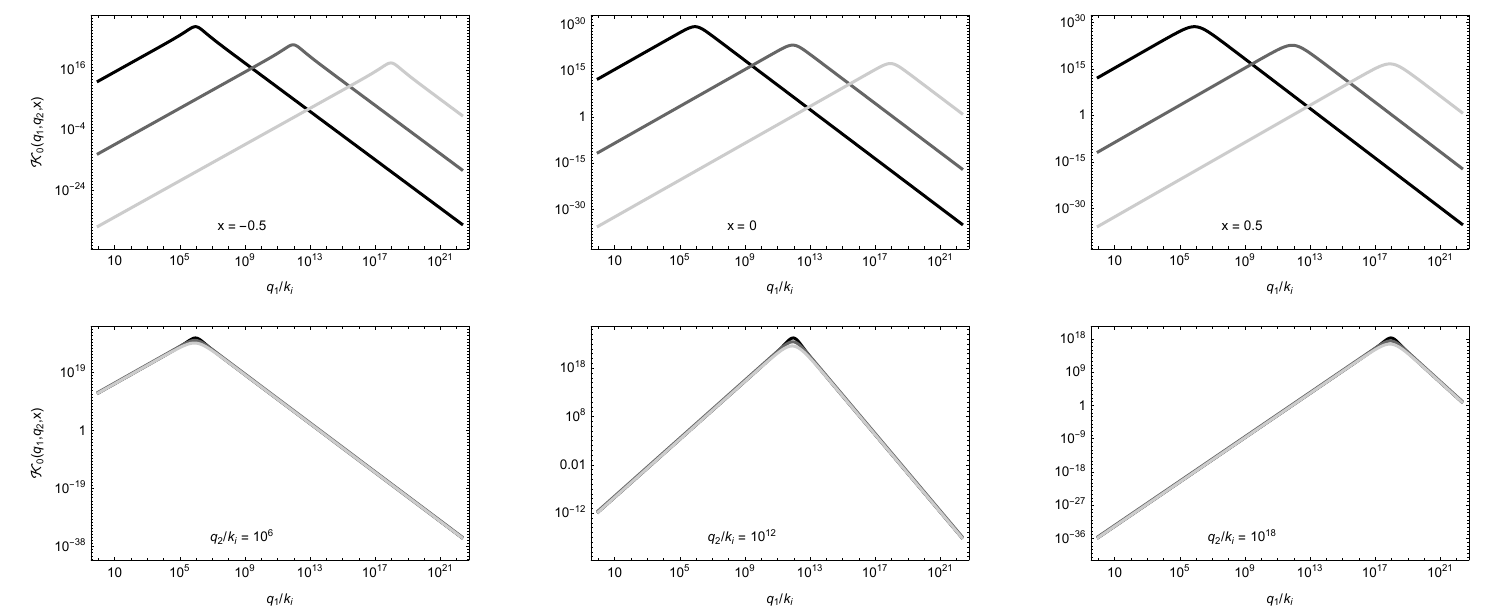}
\caption{All the plots are made using a log-log scale. Everywhere, the $y-$axis is rescaled by a factor of $10^{130}$. \emph{First line.} The plots show the function $\mathcal{K}_0(q_1,q_2,x)$ obtained using super-Hubble modes vs $\frac{q_1}{k_\text{i}}$ for some values of $\frac{q_2}{k_\text{i}}$, keeping $x$ fixed. We select $\frac{q_2}{k_\text{i}}=10^6$ (black lines), $\frac{q_2}{k_\text{i}}=10^{12}$ (dark gray lines) and $\frac{q_2}{k_\text{i}}=10^{18}$ (light gray lines). The value of $x$ changes according to what is written in the plots. \emph{Second line.} The plots show the function $\mathcal{K}_0(q_1,q_2,x)$ vs $\frac{q_1}{k_\text{i}}$ obtained using super-Hubble modes for some values of $x$, keeping $\frac{q_2}{k_\text{i}}$ fixed. We select $x=-0.5$ (black lines), $x=0$ (dark gray lines) and $x=0.5$ (light gray lines). The value of $\frac{q_2}{k_\text{i}}$ changes according to what is written in the plots.}
\label{fig_K0-SH-scales}
\end{figure*}

We notice that $\mathcal{K}_0(q_1,q_2,p)$ is symmetric under the exchange $q_1\leftrightarrow q_2$, as a consequence of both the homogeneity of the spacetime and the fact of having identical scalar particles. In Fig. \ref{fig_K0-SH-scales}, we show the behavior of the function $\mathcal{K}_0$ with respect to both the momenta of the scalar particles and their emission angle. In particular, fixing the angle $x$ and one of the two momenta, say $q_2$, the kernel plotted with respect to $q_1$ reaches a maximum at $q_1=q_2$. This means that particles are more likely to be produced with the same energy, reflecting the homogeneity of the spacetime and the symmetry under the exchange of the two scalar particles. On the other hand, for a fixed $q_2$, the kernel turns out to have higher peaks for negative $x$. This shows that, on average, particles are more likely to be produced with antiparallel momenta.

Finally, we compute explicitly the comoving number density of particles. As previously remarked, doing it by using either Eq. (\ref{eq_n-continuum-final}) or Eq. (\ref{eq_n-K0}) yields the same result, of course. In particular, we obtain
\begin{equation}
na^3\simeq 2.16\times 10^{-70}\text{ GeV}^3.
\label{eq_n-calculation-comoving}
\end{equation}
Finally, we compute the number density of produced particles at the end of inflation, obtained by considering $a_{\text{f}}\equiv a(\eta_{\text{f}})\simeq 2.68\times 10^{-32}$:
\begin{equation}
n\simeq 1.12\times 10^{25}\text{ GeV}^3.
\label{eq_n-calculation-noncomoving}
\end{equation}
This is the physical number density of particles produced from the process $h\to\phi\phi$ within a restricted range of super-Hubble graviton momenta, during the last $\sim 4$ e-foldings of the inflation. In the calculation, we only consider the scalar modes that are super-Hubble in this time interval. We remark that the super-Hubble approximation is not only a mathematical simplification, but rather it aims to enhance the predictability of our model with respect to the classical particle background, that is what we expect to detect in a classical experiment.

\begin{figure}
\includegraphics[width=0.48\textwidth]{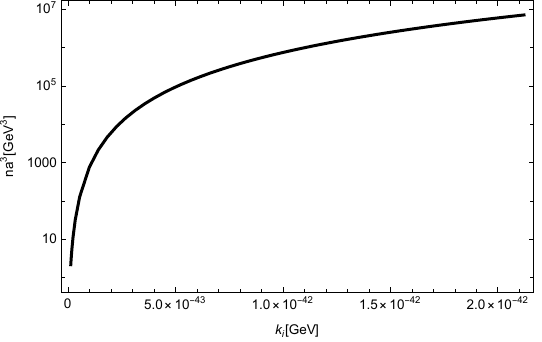}
\caption{Comoving number density of produced particles as a function of the energy scale $k_\text{i}$. The plot is made using a lin-log scale. The $y-$axis is rescaled by a factor of $10^{-70}$.}
\label{fig_nVSki}
\end{figure}

The results in Eqs. (\ref{eq_n-calculation-comoving}) and (\ref{eq_n-calculation-noncomoving}) are obtained within a very specific choice of the parameter $N_\text{piv}$, indicating the number of e-foldings after which the pivot scale exits the horizon. In particular, we select $N_\text{piv}=10$, while, more in general, as adopted by Planck mission \cite{Planck:2018jri}, its value ranges as $N_\text{piv}\in[5,10]$. We study the dependence of the number density of particles on $N_\text{piv}$. Specifically, keeping the pivot scale $k_\text{piv}$ fixed, we have $k_\text{i}=k_\text{piv}e^{-N_\text{piv}}$, and therefore varying $N_\text{piv}$ means varying $k_\text{i}$, that is, assuming a different inflationary energy scale. In Fig. \ref{fig_nVSki}, we plot the comoving number density of produced particles as a function of $k_\text{i}$. We find that $na^3\propto k_\text{i}^3$, and accordingly the physical number density of particles $n$ turns out to be constant with respect to $k_\text{i}$.

\section{Entanglement generation}{\label{sec:entanglement}}

We have studied the production of scalar particles from gravitons during the inflationary stage. At first order, an incoming one graviton state evolves into a coherent superposition of the one graviton and two scalar sectors. This is \emph{de facto} a coherent superposition of momentum channels, and therefore the resulting state can exhibit quantum entanglement in momentum space \cite{Ball:2005xa, Fuentes:2010dt, Martin-Martinez:2012chf}. This makes the model sensitive to how correlations are distributed in momentum, which particle counting alone cannot reveal.

In particular, within the inflationary scenario, the comoving Hubble radius $r_H\equiv\frac{1}{aH_I}$ naturally divides the total Hilbert space of states into sub- and super-Hubble subspaces \cite{PhysRevD.86.045014, Kiefer:1998qe}. Hence, during the particle production process, quantum correlations can be generated between scalar modes lying on opposite sides of the Hubble scale. Thus, the entanglement analysis looks at these specific modes, considering momenta within a narrow interval around the Hubble horizon. Arbitrarily, we assign ${\bf q_1}$ to the super-Hubble scalar sector and ${\bf q_2}$ to the sub-Hubble sector. We retain the reference scale $k_*$ as the Hubble horizon, and we thus have $q_1<k_*<q_2$.

Momentum conservation, in this case, requires the graviton momentum $p$ to take values below and above the horizon. In general, we may consider $\alpha k_*<p<\beta k_*$, with $\alpha<1$ and $\beta>1$. For simplicity, we set $\alpha k_*$ as the lower bound for $q_1$, and $\beta k_*$ as the upper bound for $q_2$. Momentum conservation further constrains $|p-q_1|<q_2<p+q_1$, yielding
\begin{equation}
\mathcal{D}_e:\begin{cases}
\alpha k_*<q_1<k_*\\ \text{max}\left\{k_*,|p-q_1|\right\}<q_2<\text{min}\left\{\beta k_*,p+q_1\right\}
\end{cases},
\label{eq_integration-domain-entanglement}
\end{equation}
that is, the analogue of $\mathcal{D}_n$ concerning the entanglement calculation. It is straightforward to check that, considering $p\in[\alpha k_*,\beta k_*]$, the integration domain $\mathcal{D}_e$ is well-defined if and only if $\alpha>\frac12$. Accordingly, we consider $\alpha=0.75$. On the other hand, at first glance $\beta$ may seemingly take any value. However, deeply sub-Hubble modes would yield UV sensitivity related to the Bunch-Davies vacuum, and therefore we should restrict the sub-Hubble interval as well. In particular, we take $\beta=1.25$.

Remarkably, within this integration domain we can check that the production process $h\to\phi\phi$ remains perturbative, that is, $\mathcal{P}_{\bf p}\ll 1$. We can thus study the generated quantum correlations within our perturbative framework.

For the entanglement calculation, the super-Hubble approximations for the modes cannot be used. Beyond the integration domain, the integral $\mathcal{I}_0$ should be recomputed using the complete mode solutions. Specifically, it reads
\begin{equation}
\mathcal{I}_0(q_1,q_2,x)=\frac{H_I}{(2q_1q_2p)^{\frac32}}\mathcal{B}(q_1,q_2,p),
\label{eq_I0-entanglement}
\end{equation}
with $\mathcal{B}(q_1,q_2,x)$ explicitly reported in Appendix \ref{app_Imassless}, and $p=\sqrt{q_1^2+q_2^2+2q_1q_2x}$. The function $\mathcal{K}_0$ is consequently modified.

\subsection{Quantum state and von Neumann entropy}{\label{subsec:von-neumann-entropy}}

We now characterize the quantum correlations generated during the process $h\rightarrow\phi\phi$. In this case, it is required to factorize the Hilbert space related to the cross-Hubble configurations as
\begin{equation}
 {\cal H}=
 {\cal H}_{h}\otimes
 {\cal H}_{\phi,\mathrm{SH}}\otimes
 {\cal H}_{\phi,\mathrm{sub}} .
\label{eq_Hilbert-factorization-entanglement}
\end{equation}
According to this factorization, the initial state can be written as
\begin{equation}
\ket{\Psi_{{\bf p},\lambda}}=\ket{h({\bf p},\lambda)}\ket{0}_\text{SH}\ket{0}_\text{sub},
\end{equation}
and its first-order evolution restricted to the selected one-pair sector is
\begin{equation}
\begin{split}
 |\Psi_{{\bf p},\lambda}\rangle
 =&
 \frac{1}
 {\sqrt{1+{\cal P}_{\bf p}^{(\lambda)}}}
 \Bigg[
 |h({\bf p},\lambda)\rangle
 |0\rangle_{\rm SH}|0\rangle_{\rm sub}
 \\
 &+
 \sum_{\substack{{\bf q}_1\in{\rm SH}\\
                  {\bf q}_2\in{\rm sub}}}
 C_{{\bf q}_1{\bf q}_2}^{(\lambda)}
 |0\rangle_h
 |\phi({\bf q}_1)\rangle_{\rm SH}
 |\phi({\bf q}_2)\rangle_{\rm sub}
 \Bigg].
\end{split}
\label{eq_total-state-entanglement}
\end{equation}
In the expression above, $C_{{\bf q_1},{\bf q_2}}^{(\lambda)}$ is the same as in Eq. (\ref{eq_C-explicit-production}), but with $\mathcal{I}_0$ coming from Eq. (\ref{eq_I0-entanglement}).

The complete state in Eq. \eqref{eq_total-state-entanglement} is pure, thus its von Neumann entropy vanishes. Conversely, a non-vanishing entanglement entropy is obtained when considering a bipartition of the system.

We start from the bipartition that separates the graviton from the complete scalar sector. Tracing over both scalar sectors yields
\begin{equation}
 \rho_h^{({\bf p},\lambda)}
 =
 \frac{
 |h({\bf p},\lambda)\rangle
 \langle h({\bf p},\lambda)|
 +
 {\cal P}_{\bf p}^{(\lambda)}
 |0\rangle_h\langle0|
 }{
 1+{\cal P}_{\bf p}^{(\lambda)}
 }.
\label{eq_rho-h-entanglement}
\end{equation}
Defining
\begin{equation}
 r_{\bf p}^{(\lambda)}
 \equiv
 \frac{
 {\cal P}_{\bf p}^{(\lambda)}
 }{
 1+{\cal P}_{\bf p}^{(\lambda)}
 },
\label{eq_r-entanglement}
\end{equation}
the graviton--scalar entropy is written as
\begin{equation}
 S_{h|\phi\phi}^{({\bf p},\lambda)}
 =
 H_2\left(r_{\bf p}^{(\lambda)}\right)\equiv
 -r_{\bf p}^{(\lambda)}\ln r_{\bf p}^{(\lambda)}-(1-r_{\bf p}^{(\lambda)})\ln(1-r_{\bf p}^{(\lambda)}),
\label{eq_graviton-scalar-entanglement}
\end{equation}
where $H_2\left(r_{\bf p}^{(\lambda)}\right)$ is the binary entropy.

Then, we consider the super-Hubble scalar sector and trace over both the graviton and the sub-Hubble scalar modes. In this regard, we introduce
\begin{equation}
 {\cal P}_{{\bf q}_1|{\bf p}}^{(\lambda)}
 \equiv
 \sum_{{\bf q}_2\in{\rm sub}}
 \left|C_{{\bf q}_1{\bf q}_2}^{(\lambda)}\right|^2 .
\label{eq_single-mode-probability-entanglement}
\end{equation}
Because of momentum conservation, only ${\bf q}_2={\bf p}-{\bf q}_1$ contributes, and consequently
\begin{equation}
 \sum_{{\bf q}_1\in{\rm SH}}
 {\cal P}_{{\bf q}_1|{\bf p}}^{(\lambda)}
 =
 {\cal P}_{\bf p}^{(\lambda)} .
\label{eq_probability-sum-entanglement}
\end{equation}
The reduced density matrix is then
\begin{equation}
\begin{split}
 \rho_{\rm SH}^{({\bf p},\lambda)}
 =&
 {\rm Tr}_{h,\rm sub}
 \left(
 |\Psi_{{\bf p},\lambda}\rangle
 \langle\Psi_{{\bf p},\lambda}|
 \right)
 \\
 =&
 \frac{1}{1+{\cal P}_{\bf p}^{(\lambda)}}
 \Bigg[
 |0\rangle\langle0|
 +
 \sum_{{\bf q}_1\in{\rm SH}}
 {\cal P}_{{\bf q}_1|{\bf p}}^{(\lambda)}
 |\phi({\bf q}_1)\rangle
 \langle\phi({\bf q}_1)|
 \Bigg],
\end{split}
\label{eq_rho-SH-entanglement}
\end{equation}
with eigenvalues
\begin{equation}
 \lambda_0=
 \frac{1}{1+{\cal P}_{\bf p}^{(\lambda)}},
 \qquad
 \lambda_{{\bf q}_1}=
 \frac{
 {\cal P}_{{\bf q}_1|{\bf p}}^{(\lambda)}
 }{
 1+{\cal P}_{\bf p}^{(\lambda)}
 }.
\label{eq_rho-eigenvalues-entanglement}
\end{equation}
The corresponding entropy reads
%
%-{\rm Tr}\left(\rho_{\rm SH}^{({\bf p},\lambda)}\ln\rho_{\rm SH}^{({\bf p},\lambda)}\right)\\=&
\begin{equation}
\begin{split}
 S_{\phi_{\rm SH}|(h+\phi_{\rm sub})}
 =&
 \ln\left(1+{\cal P}_{\bf p}^{(\lambda)}\right)
 \\&-
 \frac{1}{1+{\cal P}_{\bf p}^{(\lambda)}}
 \sum_{{\bf q}_1\in{\rm SH}}
 {\cal P}_{{\bf q}_1|{\bf p}}^{(\lambda)}
 \ln
 {\cal P}_{{\bf q}_1|{\bf p}}^{(\lambda)},
\end{split}
\label{eq_entropy-total-exact}
\end{equation}
and measures the bipartite entanglement $\phi_{\rm SH}\;\big|\;
 \left(h+\phi_{\rm sub}\right)$.

Equivalently, if we trace over the graviton and the super-Hubble scalar sector, the reduced density matrix of the sub-Hubble scalar modes has the same non-vanishing spectrum as Eq. \eqref{eq_rho-SH-entanglement}, because momentum conservation fixes ${\bf q}_2={\bf p}-{\bf q}_1$. Explicitly, it yields
\begin{equation}
 S_{\phi_{\rm SH}|(h+\phi_{\rm sub})}
 =
 S_{\phi_{\rm sub}|(h+\phi_{\rm SH})}
 \equiv S_{\rm ent}.
\label{eq_scalar-partitions-equality}
\end{equation}
This equality is a consequence of the structure of the $h\rightarrow\phi\phi$ process and is not a generic property of a generic tripartite state.

In the perturbative regime ${\cal P}_{\bf p}^{(\lambda)}\ll1$, so Eq. \eqref{eq_entropy-total-exact} reduces to
\begin{equation}
 S_{\phi_{\rm SH}|(h+\phi_{\rm sub})}
 \simeq
 -\sum_{{\bf q}_1\in{\rm SH}}
 {\cal P}_{{\bf q}_1|{\bf p}}^{(\lambda)}
 \left[
 \ln
 {\cal P}_{{\bf q}_1|{\bf p}}^{(\lambda)}
 -1
 \right].
\label{eq_entropy-total-pert}
\end{equation}
We can now recover the continuum entropy density. We define the comoving entropy density by summing the perturbative contribution over the incoming graviton modes and over the two physical helicities:
\begin{equation}
 sa^3\equiv
 \frac{1}{V}
 \sum_{\lambda=\pm2}
 \sum_{\bf p}
 S_{\phi_{\rm SH}|(h+\phi_{\rm sub})} .
\label{eq_entropy-density-discrete}
\end{equation}
In order to transition to the continuum, the probability of a single discrete momentum state is replaced by a probability density within the following prescription \cite{cover2006elements}.
\begin{itemize}
\item[-] In Eq. (\ref{eq_entropy-total-pert}), the probability of a single discrete momentum state is given by Eq. (\ref{eq_single-mode-probability-entanglement}). In the continuum limit, we can keep the same expression for the probability outside the logarithm.
\item[-] Conversely, in order to consistently transition from discrete to continuum, the probability inside the logarithm should be replaced according to
\begin{equation}
\ln
 {\cal P}_{{\bf q}_1|{\bf p}}^{(\lambda)}\to \ln{\cal W_\lambda}(q_1,q_2,x),
\end{equation}
with
\begin{align}
 {\cal W_\lambda}(q_1,q_2,x)\equiv&
 |C_{{\bf q}_1{\bf q}_2}^{(\lambda)}|^2
 \frac{Vk_{\rm i}^3}{(2\pi)^3}.
\end{align}
\item[-] As in the calculation of the particle number, we assume $|C_{{\bf q}_1{\bf q}_2}^{(-2)}|^2=|C_{{\bf q}_1{\bf q}_2}^{(+2)}|^2$, that is, the two physical graviton polarizations are equally probable. Accordingly, by using Eq. (\ref{eq_C-squared-production}) and Eq. (\ref{eq_K-function}), we get
\begin{equation}
{\cal W_\lambda}(q_1,q_2,x)=\frac{k_\text{i}^4}{8\pi^3}\left(\frac{k_{\text{i}}}{\overline{M}_{\text{Pl}}}\right)^2\frac{\mathcal{K}_0(q_1,q_2,x)}{q_1^2q_2^2}.
\end{equation}
\end{itemize}
Hence, Eq. (\ref{eq_entropy-density-discrete}) becomes
\begin{equation}
sa^3=-\frac{1}{V}\sum_{\bf p}\sum_{{\bf q_1}\in \text{SH}}\sum_{{\bf q_2}\in \text{sub}}\sum_{\lambda=\pm 2}|C_{{\bf q}_1{\bf q}_2}^{(\lambda)}|^2\left[
 \ln
 {\cal W_\lambda}
 -1
 \right].
\end{equation}
In the continuum limit, we obtain
\begin{equation}
\begin{split}
&sa^3=-\frac{k_\text{i}^3}{4\pi^4\overline{M}^2_\text{Pl}}\int_{\tilde{\mathcal{D}}_e} dq_1dq_2\int_{x_-(q_1,q_2)}^{x_+(q_1,q_2)} dx \mathcal{K}_0(q_1,q_2,x)\times \\
&\qquad \qquad \qquad \qquad \quad \times \left[\ln\mathcal{W}(q_1,q_2,x)-1\right],
\end{split}
\end{equation}
where we define
\begin{subequations}
\begin{align}
&x_-(q_1,q_2)\equiv \text{max}\left\{-1,\frac{(\alpha k_*)^2-q_1^2-q_2^2}{2q_1q_2}\right\},
\label{eq_xmin-ent}\\
&x_+(q_1,q_2)\equiv\text{min}\left\{1,\frac{(\beta k_*)^2-q_1^2-q_2^2}{2q_1q_2}\right\}
\label{eq_xmax-ent}.
\end{align}
\end{subequations}
Notably, these are different from the bounds for $x$ used for the calculation of the particle number, as there we consider a different range of graviton momenta.

The integration domain $\tilde{\mathcal{D}}_e$ is the intersection between the rectangle $[\alpha k_*,k_*]\times [k_*,\beta k_*]$ and the constraints to guarantee that $x_-<x_+$ independently of $q_1$ and $q_2$. However, this condition turns out to be satisfied automatically within the rectangle, and therefore we simply have
\begin{equation}
\tilde{\mathcal{D}}_e:\begin{cases}
\alpha k_*<q_1<k_*\\ k_*<q_2<\beta k_*
\end{cases}.
\label{eq_D-ent-tilde}
\end{equation}
In the calculation of the entanglement, we are mainly interested in the final particle states, as the latter show quantum correlations in momentum space. Accordingly, we define
\begin{equation}
\sigma(q_1,q_2,x)\equiv -\mathcal{K}_0(q_1,q_2,x)\left[\ln\mathcal{W}(q_1,q_2,x)-1\right].
\label{eq_sigma-entanglement}
\end{equation}
The function $\sigma(q_1,q_2,x)$ represents \emph{de facto} the comoving entanglement entropy spectrum concerning the final particle states. It characterizes the momentum-space distribution of the quantum correlations generated between the scalar particle pairs from the process $h\to \phi\phi$. In this respect, the entanglement analysis turns out to be a probe of the structure of the aforementioned process, rather than a mere side effect.

\begin{figure*}
\includegraphics[width=1.0\textwidth]{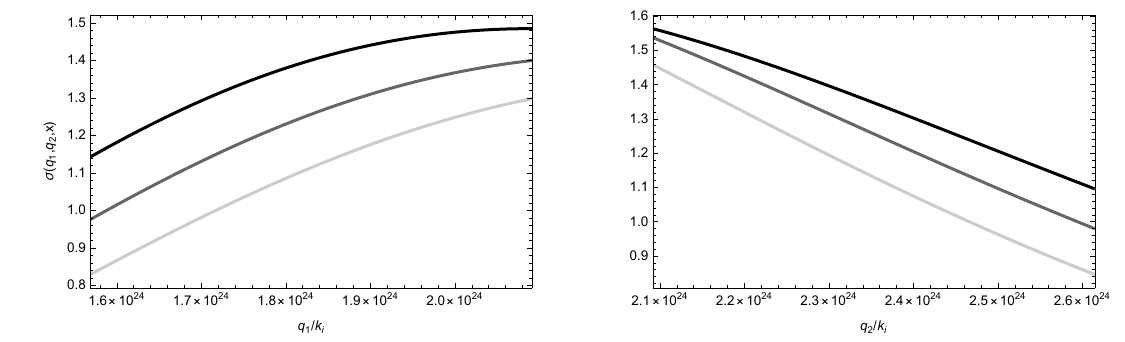}
\caption{Integrand function $\sigma(q_1,q_2,x)$ evaluated at $x=-0.5$. Both the plots are made using a lin-lin scale. Everywhere, the $y-$axis is rescaled by a factor of $10^{144}$. \emph{Left plot.} The dependence of $\sigma$ on the rescaled super-Hubble momentum variable $\frac{q_1}{k_\text{i}}$ is shown for $\frac{q_2}{k_\text{i}}=\left(1+\frac{\beta-1}{4}\right)e^{N_*}$ (black line), $\frac{q_2}{k_\text{i}}=\left(1+\frac{\beta-1}{2}\right)e^{N_*}$ (dark gray line) and $\frac{q_2}{k_\text{i}}=\left(1+\frac{3(\beta-1)}{4}\right)e^{N_*}$ (light gray line). \emph{Right plot.} The dependence of $\sigma$ on the sub-Hubble momentum variable $\frac{q_2}{k_\text{i}}$ is shown for $\frac{q_1}{k_\text{i}}=\left(\alpha+\frac{3(1-\alpha)}{4}\right)e^{N_*}$ (black line), $\frac{q_1}{k_\text{i}}=\left(\alpha+\frac{1-\alpha}{2}\right)e^{N_*}$ (dark gray line) and $\frac{q_1}{k_\text{i}}=\left(\alpha+\frac{1-\alpha}{4}\right)e^{N_*}$ (light gray line).}
\label{fig_sigmaVSyz}
\end{figure*}

\begin{figure}
\includegraphics[width=0.5\textwidth]{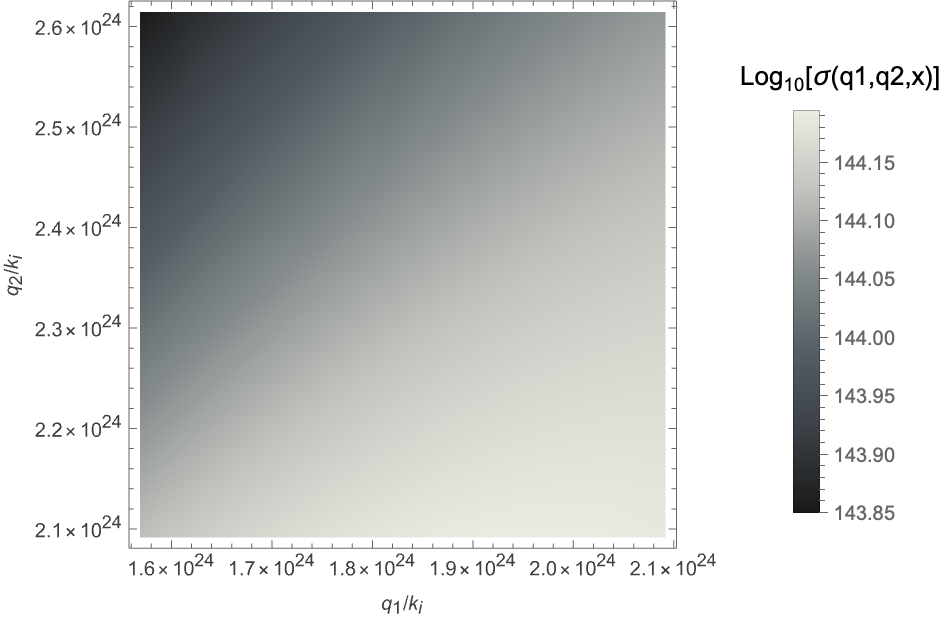}
\caption{Density plot of the integrand function $\sigma(q_1,q_2,x)$ in the $\left(\frac{q_1}{k_\text{i}}-\frac{q_2}{k_\text{i}}\right)$ plane, evaluated at $x=-0.5$.}
\label{fig_sigmadensity}
\end{figure}

As shown in Fig. \ref{fig_sigmaVSyz}, the entropy-density integrand increases as the super-Hubble momentum approaches $k_*$ from below and the sub-Hubble momentum approaches it from above. This indicates that scalar configurations close to the reference Hubble scale provide the dominant contribution within the selected domain. The same behavior is illustrated by the two-dimensional distribution in Fig. \ref{fig_sigmadensity}.

\begin{figure}
\includegraphics[width=0.48\textwidth]{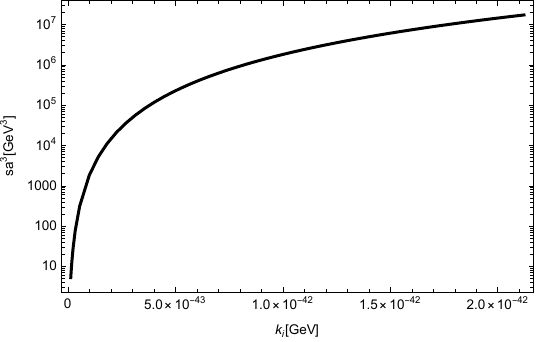}
\caption{Comoving von Neumann entropy density as a function of the energy scale $k_\text{i}$. The plot is made using a lin-log scale. The $y-$axis is rescaled by a factor of $10^{-68}$.}
\label{fig_sVSki}
\end{figure}

Finally, as we have done for the number density of particles, we want to relate the entanglement entropy to the inflationary energy scale $k_\text{i}$. For this purpose, we investigate the dependence of the comoving von Neumann entropy on this scale. We show the corresponding plot in Fig. \ref{fig_sVSki}. In analogy to what we obtain for the number density of particles, we find that $sa^3\propto k_\text{i}^3$. Specifically, for $k_\text{i}\simeq 1.43\times 10^{-44}\text{ GeV}$, that is, $N_\text{piv}=10$, we find $sa^3\simeq 5.26\times 10^{-68}\text{ GeV}^3$.

\subsection{Entanglement between the scalar sectors}{\label{subsec_entScalar}}

%The von Neumann entropy computed in the previous subsection characterizes the entanglement of a given scalar sector with its complementary subsystem, which also includes the graviton\footnote{We emphasize that Figs. \ref{fig_sigmaVSyz} and \ref{fig_sigmadensity} refer to the entropy associated with the bipartition $\phi_{\rm SH}|(h+\phi_{\rm sub})$.}. Consequently, it does not directly measure the entanglement between the two scalar sectors. To isolate these correlations, we trace over the graviton in Eq.~\eqref{eq_total-state-entanglement} and compute the logarithmic negativity of the resulting mixed scalar state \cite{VidalWerner2002, PlenioVirmani2007}.

The von Neumann entropy computed in Sect. \ref{subsec:von-neumann-entropy} quantifies the entanglement related to the bipartition $\phi_{\rm SH}|(h+\phi_{\rm sub})$, rather than directly between the two scalar sectors. To isolate the latter, we trace out the graviton and evaluate the logarithmic negativity of the resulting mixed scalar state \cite{VidalWerner2002, PlenioVirmani2007}.

We consider a fixed incoming graviton momentum $\mathbf p$ and helicity $\lambda$. For the scalar particle states, throughout this subsection, the sum over ${\bf q_1}$ and ${\bf q_2}$ is restricted to $\tilde{\mathcal{D}}_e$, with the emission angle constrained between $x_-$ and $x_+$, defined in Eqs. (\ref{eq_xmin-ent}) and (\ref{eq_xmax-ent}), respectively. Accordingly, all the relevant quantities, such as the quantum state $\ket{\Psi_{{\bf p},\lambda}}$ and the transition probability $\mathcal{P}_{\bf p}^{(\lambda)}$, are those of Sect. \ref{subsec:von-neumann-entropy}.

The reduced scalar density operator, traced over the graviton Hilbert space using orthogonality relations between the states, reads
\begin{align}
\rho_{\phi\phi}^{(p,\lambda)}
\equiv&\operatorname{Tr}_h\bigl(
|\Psi_{p,\lambda}\rangle\langle\Psi_{p,\lambda}|\bigr)
\nonumber\\
=&\frac{1}{1+\mathcal P_{\bf p}^{(\lambda)}}\bigg[
|0\rangle_{\rm SH}|0\rangle_{\rm sub}
{}_{\rm SH}\langle0|{}_{\rm sub}\langle0|
\nonumber\\
&+\sum_{{\bf q_1},{\bf q_2},{\bf q_1'},{\bf q_2'}}
C_{{\bf q_1}{\bf q_2}}^{(\lambda)}
C_{{\bf q_1'}{\bf q_2'}}^{(\lambda)*}
%\nonumber\\
%&\qquad\times
|\phi({\bf q_1})\rangle_{\rm SH}
|\phi({\bf q_2})\rangle_{\rm sub}
\nonumber\\
&\times{}_{\rm SH}\langle\phi({\bf q_1'})|
{}_{\rm sub}\langle\phi({\bf q_2'})|\bigg].
\label{eq:logneg_phi_rho}
\end{align}
Notably, it is straightforward to verify that this operator is Hermitian and has unit trace.
Further, since $\mathcal P_{\bf p}^{(\lambda)}>0$, we also find that the scalar state is mixed, indeed
\begin{align}
\operatorname{Tr}\bigl[(\rho_{\phi\phi}^{(p,\lambda)})^2\bigr]
&=\frac{1+(\mathcal P_{\bf p}^{(\lambda)})^2}
{(1+\mathcal P_{\bf p}^{(\lambda)})^2}<1.
\label{eq:logneg_phi_purity}
\end{align}
The logarithmic negativity is defined as the logarithm of the trace norm of its partially transposed density operator
\begin{equation}
\mathcal E_{\mathcal N}^{(\lambda)}(p)
\equiv\ln\left\|
\left(\rho_{\phi\phi}^{(p,\lambda)}\right)^{T_{\rm sub}}
\right\|_1,
\qquad
\|X\|_1\equiv\operatorname{Tr}\sqrt{X^\dagger X}.
\label{eq:logneg_phi_def}
\end{equation}
In particular, the partial transpose with respect to the sub-Hubble sector exchanges the sub-Hubble indices between ket and bra, while it leaves the super-Hubble indices unchanged. On the other hand, the vacuum projector is unaltered, thus, we obtain 
\begin{align}
\left(\rho_{\phi\phi}^{(p,\lambda)}\right)^{T_{\rm sub}}
=&\frac{1}{1+\mathcal P_{\bf p}^{(\lambda)}}\bigg[
|0\rangle_{\rm SH}|0\rangle_{\rm sub}
{}_{\rm SH}\langle0|{}_{\rm sub}\langle0|
\nonumber\\
&+\sum_{{\bf q_1},{\bf q_2},{\bf q_1'},{\bf q_2'}}
C_{{\bf q_1}{\bf q_2}}^{(\lambda)}C_{{\bf q_1'}{\bf q_2'}}^{(\lambda)*}\,|\phi({\bf q_1})\rangle_{\rm SH}
|\phi({\bf q_2'})\rangle_{\rm sub}\notag\\
&\times{}_{\rm SH}\langle\phi({\bf q_1'})|
{}_{\rm sub}\langle\phi({\bf q_2})|\bigg].
\label{eq:logneg_phi_pt}
\end{align}
Since partial transpose preserves the operator's hermiticity, in order to explicitly write Eq. \eqref{eq:logneg_phi_def}, we proceed by computing its eigenvalues.
The scalar vacuum and the product states associated with the diagonal momentum channels, with ${\bf q_1}={\bf q_1'}$ and consequently ${\bf q_2}={\bf q_2'}$ for fixed $\mathbf p$, are eigenvectors with eigenvalues
\begin{subequations}
\begin{align}
&\Lambda_0=\frac{1}{1+\mathcal P_{\bf p}^{(\lambda)}},
\label{eq:logneg_phi_lambda0}\\
&\Lambda_{{\bf q_1}{\bf q_2}}=
\frac{|C_{{\bf q_1}{\bf q_2}}^{(\lambda)}|^2}
{1+\mathcal P_{\bf p}^{(\lambda)}},
\label{eq:logneg_phi_lambdadiag}
\end{align}
\end{subequations}
respectively. The remaining contributions to the partially transposed operator, given by the two distinct allowed channels $({\bf q_1},{\bf q_2})$ and $({\bf q_1'},{\bf q_2'})$, with ${\bf q_1}\neq {\bf q_1'}$, act on the two-dimensional subspace spanned by
\begin{equation}
\bigl\{
|\phi({\bf q_1})\rangle_{\rm SH}|\phi({\bf q_2'})\rangle_{\rm sub},\;
|\phi({\bf q_1'})\rangle_{\rm SH}|\phi({\bf q_2})\rangle_{\rm sub}
\bigr\}.
\label{eq:logneg_phi_basis}
\end{equation}
By diagonalizing these contributions, each unordered pair of distinct channels provides a positive and a negative eigenvalue
\begin{equation}
\Lambda_{{\bf q_1}{\bf q_1'}}^{\pm}
=\pm\frac{|C_{{\bf q_1}{\bf q_2}}^{(\lambda)}
C_{{\bf q_1'}{\bf q_2'}}^{(\lambda)}|}
{1+\mathcal P_{\bf p}^{(\lambda)}}.
\label{eq:logneg_phi_lambdapair}
\end{equation}
Here, since any diagonalized Hermitian operator can be written as $X=U\operatorname{diag}(\Lambda_j)U^\dagger$ with $U$ a unitary matrix, it follows
\begin{align}
\left\|\left(\rho_{\phi\phi}^{(p,\lambda)}\right)^{T_{\rm sub}}\right\|_1
&=\frac{1+
\left(\displaystyle\sum_{{\bf q_1},{\bf q_2}}|C_{{\bf q_1}{\bf q_2}}^{(\lambda)}|\right)^2}
{1+\mathcal P_{\bf p}^{(\lambda)}}.
\label{eq:logneg_phi_tracenorm}
\end{align}

Finally, the logarithmic negativity of the mixed scalar state reads
\begin{equation}
\mathcal E_{\mathcal N}^{(\lambda)}(p)
=\ln\left[
\frac{1+\left(\sum_{{\bf q_1},{\bf q_2}}
|C_{{\bf q_1}{\bf q_2}}^{(\lambda)}|\right)^2}
{1+\mathcal P_{\bf p}^{(\lambda)}}
\right].
\label{eq:logneg_phi_result}
\end{equation}
Unlike the von Neumann entropy previously considered, the result in Eq.~\eqref{eq:logneg_phi_result} provides a direct measure of the quantum correlations between the super-Hubble and sub-Hubble scalar sectors, after tracing out the graviton degrees of freedom. Indeed, by using the definition of $\mathcal P_{\bf p}^{(\lambda)}$, Eq.~\eqref{eq:logneg_phi_result} can be equivalently expressed as
\begin{equation}
\mathcal E_{\mathcal N}^{(\lambda)}(p)
=
\ln\left[
1+\frac{2}{1+\mathcal P_{\bf p}^{(\lambda)}}
\sum_{i<j}|C_i^{(\lambda)}C_j^{(\lambda)}|
\right],
\label{eq:logneg_phi_channels}
\end{equation}
where $i$ and $j$ denote distinct momentum channels, $({\bf q_1},{\bf q_2})$ and $({\bf q_1'},{\bf q_2'})$ compatible with momentum conservation. Consequently, a positive logarithmic negativity requires at least two contributing momentum channels, whose coherent superposition generates non-separable correlations.
In other words, a non-vanishing particle production probability does not necessarily imply entanglement between the two scalar sectors. 

To investigate the continuum limit, we proceed in analogy to quantum field theory approaches by considering the infinite volume behavior of Eq. \eqref{eq:logneg_phi_result}. For a fixed incoming graviton momentum and helicity, the finite-volume normalization implies that $\mathcal P_{\bf p}^{(\lambda)}$ approaches a finite value, whereas $\left(\sum_{{\bf q_1},{\bf q_2}}|C_{{\bf q_1}{\bf q_2}}^{(\lambda)}|\right)^2$ grows linearly with $V$. We introduce two related prescriptions to address this divergent behavior.

First, choosing a reference graviton momentum $p'$, we introduce the difference of logarithmic negativity with respect to this scale, according to
\begin{equation}
\Delta\mathcal E_{\mathcal N}^{(p',\lambda)}(p)
\equiv\lim_{V\rightarrow\infty}\left(\mathcal E_{\mathcal N}^{(\lambda)}(p)-\mathcal E_{\mathcal N}^{(\lambda)}(p')\right).
\label{eq_delta-neg-def}
\end{equation}
Explicitly, considering $\mathcal{P}_{\bf p}\ll 1$, we obtain
\begin{equation}
\Delta\mathcal E_{\mathcal N}^{(p',\lambda)}(p)\!=\!2\ln\!\!\left[
\frac{p'\int_{\mathcal D_e(p)}dq_1dq_2\sqrt{\mathcal{K}_0(q_1,q_2,x_p(q_1,q_2))}}
{p\int_{\mathcal D_e(p')}dq_1dq_2\sqrt{\mathcal{K}_0(q_1,q_2,x_{p'}(q_1,q_2))}}
\right].
\label{eq_delta-neg-integral}
\end{equation}
where the prime denotes quantities related to the scale $p'$. Moreover, the integration domain $\mathcal{D}_e(p)$ is defined in Eq. (\ref{eq_integration-domain-entanglement}), and $x_p(q_1,q_2)\equiv \frac{p^2-q_1^2-q_2^2}{2q_1q_2}$.

It is straightforward to check that the volume dependence cancels, allowing us to characterize the relative variation of scalar entanglement.

\begin{figure}
\includegraphics[width=0.48\textwidth]{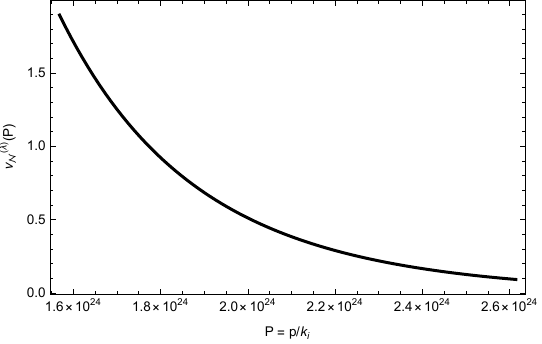}
\caption{Negativity density as a function of the rescaled graviton momentum $P=\frac{p}{k_\text{i}}$. The plot is made using a lin-lin scale. The $y$-axis is shown in units of $10^{-71}\text{ GeV}^3$.}
\label{fig_nu}
\end{figure}

Alternatively, to infer more information in the continuum limit, we can consider ordinary negativity $\mathcal N^{(\lambda)}_{\bf p}$, defined by \cite{VidalWerner2002, PlenioVirmani2007}
\begin{equation}
    \mathcal E_{\mathcal N}^{(\lambda)}(p)
=\ln\left(2{\mathcal N_{\bf p}^{(\lambda)}}+1\right).
\label{eq:logneg_neg}
\end{equation}
Explicitly, the negativity reads
\begin{equation}
{\mathcal N_{\bf p}^{(\lambda)}}
=
\frac{\left(\sum_{{\bf q_1},{\bf q_2}}
|C_{{\bf q_1}{\bf q_2}}^{(\lambda)}|\right)^2-\mathcal P_{\bf p}^{(\lambda)}}
{2\left(1+\mathcal P_{\bf p}^{(\lambda)}\right)}.
\label{eq:neg}
\end{equation}
By considering the continuum limit, that is, $V\to \infty$, any volume dependence can be removed if considering the density negativity:
\begin{equation}
\nu_\mathcal{N}^{(\lambda)}(p)\equiv\lim_{V\to \infty}\frac{\mathcal{N}_{\bf p}^{(\lambda)}}{V}=\frac{\left(\sum_{{\bf q_1},{\bf q_2}}
|C_{{\bf q_1}{\bf q_2}}^{(\lambda)}|\right)^2}
{2V\left(1+\mathcal P_{\bf p}^{(\lambda)}\right)}.
\label{eq_neg-density-def}
\end{equation}
%
%Remarkably, in this limit we neglect $\mathcal{P}_{\bf p}^{(\lambda)}$ because it is constant in $V$, while the other term is proportional to it. 
In the infinite-volume limit, the term proportional to $\mathcal{P}_{\bf p}^{(\lambda)}$ vanishes after dividing by $V$, whereas the squared sum of transition amplitudes yields a finite contribution.
Explicitly, in the perturbative regime, we obtain
\begin{equation}
\nu_\mathcal{N}^{(\lambda)}(p)\!\simeq\!\frac{k_\text{i}^3}
{32\pi^4M_{\rm Pl}^2p^2}\!\left[\int_{\mathcal D_e(p)}\!\!\!\!dq_1dq_2
\sqrt{\mathcal{K}_0(q_1,q_2,x_p(q_1,q_2)}\right]^2.
\label{eq_neg-density-integral}
\end{equation}
Notably, the two prescriptions are directly related through
\begin{equation}
\Delta\mathcal E_{\mathcal N}^{(p',\lambda)}(p)
=
\ln\left[
\frac{\nu_{\mathcal N}^{(\lambda)}(p)}
{\nu_{\mathcal N}^{(\lambda)}(p')}
\right].
\label{eq:logneg_density_relation}
\end{equation}
In Fig. \ref{fig_nu}, we plot the negativity density $\nu_\mathcal{N}^{(\lambda)}$ as a function of the rescaled graviton momentum $P=\frac{p}{k_\text{i}}$. We remark that the behavior of $\Delta\mathcal{E}_\mathcal{N}^{(p',\lambda)}$ is related to $\nu_\mathcal{N}^{(\lambda)}$ by Eq. (\ref{eq:logneg_density_relation}), and therefore we do not plot it separately.

The negativity density decreases with increasing incoming graviton momentum, showing a trend similar to that of the transition probability. Within the selected momentum domain, this suggests that lower momentum gravitons contribute more strongly to the scalar sector quantum correlations.

\section{Conclusions}{\label{sec:conclusions}}

In this work, we investigated the generation of quantum entanglement between scalar field modes induced by quantized tensor perturbations of the metric during de Sitter inflation. The cosmological background was treated classically, while both the tensor perturbations and the scalar field were quantized.

As a first step, we considered the expansion of the Einstein-Hilbert action around the Minkowski metric. At second-order in the metric perturbation, we obtained the consistent quadratic Lagrangian describing the propagation of a massless spin-2 tensor field in Minkowski spacetime. Then, we showed that a real scalar field minimally coupled to gravity generates an interaction between the metric perturbation and the matter field itself in the linearized theory. In particular, we proved that the interaction can be treated perturbatively with respect to the free terms. The overall Lagrangian described a massless spin-2 tensor field perturbatively coupled to a real scalar field in flat spacetime.

Then, we generalized the theory to a conformally flat de Sitter spacetime. Such a background is an Einstein spacetime, and therefore the flat-to-curved generalization can be done consistently by adding a further term proportional to the scalar curvature and quadratic in the metric perturbations. Overall, we were left with a Lagrangian density describing the propagation of a massless spin-2 tensor field interacting with a real scalar field in de Sitter spacetime. There, tensor field quantization proceeded straightforwardly as long as the interaction term could be treated perturbatively. We referred to the quanta of the spin-2 field as the gravitons, and they naturally turned out to be massless spin-2 particles.

The theory enjoyed gauge symmetry, restricting the number of physical propagating degrees of freedom to two. We fixed the gauge to recover the transversality and tracelessness conditions (TT gauge). In this way, the equations of motion simplified, and we solved them using an SVT decomposition of the tensor field. Then, we used the remaining gauge freedom in order to eliminate the scalar and vector modes, as the latter do not propagate, leaving the two transverse-traceless modes as the only physical degrees of freedom.

Within this framework, we studied the process $h\rightarrow\phi\phi$ perturbatively, considering free quantum fields and expanding the $\mathcal{S}-$matrix up to first order in the interaction Lagrangian. Remarkably, the tensor metric perturbation did not act as a classical external source, but rather its quanta were directly involved in the process. Accordingly, this analysis may provide a quadratic quantum correction to the Parker effect. In fact, we proved that the perturbative regime was preserved only by selecting a limited graviton momentum range, so as to keep the transition probability sufficiently small. Moreover, for the particle production estimate we limit the analysis to a super-Hubble momentum domain. From a quantum field theory perspective, this played the role of the UV cutoff, realizing \emph{de facto} a regularization prescription. In fact, we discussed that limiting the analysis to super-Hubble scales also improved the predictability of the model with respect to possible measurements of the classical particle background. 

We characterized the transition in terms of both incoming graviton momenta and outgoing scalar configurations.
In particular, its amplitude turned out to decrease with the incoming graviton momentum, while it was enhanced when the two scalars are produced in antiparallel directions and with the same momentum magnitude. 
Finally, we related our results to the inflationary energy scale, improving the predictability of our model with respect to Planck data.

Then, we investigated the quantum entanglement associated with the produced state by separating the scalar momentum space with respect to a reference Hubble scale. The resulting correlations provide a theoretical probe of the quantum dynamics of inflationary perturbations. In particular, the relation between these correlations and observable signatures in cosmological data, such as in the CMB analyses and high-redshift probes \cite{Muccino:2022rnd}, remains to be deepened in future works. 

Since the complete first-order state contained both a graviton component and a two-scalar component, the system presented a tripartite structure. This allowed us to distinguish between the entanglement of one scalar sector with the remaining degrees of freedom and the entanglement between the two scalar sectors. Specifically, we studied the former through the von Neumann entropy density and the latter considering the negativity.

We proved that, within the selected cross-Hubble momentum domain, the entropy density spectrum became larger as the scalar momenta approached the reference Hubble scale from opposite regimes, and decreased as they moved away from it. We also investigated how the comoving von Neumann entropy density depends on the initial comoving scale $k_\text{i}$, providing an additional characterization of the model's parameter dependence useful for future observational studies. 

Finally, we investigated the entanglement between the two scalar sectors considering the logarithmic negativity with respect to a fixed scale, and, equivalently, the density negativity. In particular, the latter turned out to decrease with the graviton momentum, following the behavior of the transition amplitude.

Future work may aim to extend the present analysis to massive spin-2 fields, for which scalar and vector degrees of freedom would contribute to both particle production and entanglement generation. In this respect, it would be intriguing to compare the contributions of the various helicities to the particle production. In addition, in such frameworks, the mass of the graviton is constrained by the Higuchi bound, and this may affect the particle production. Further extensions could include different kinds of matter fields, such as fermion fields or charged scalars, different cosmological backgrounds, e.g. quasi-de Sitter spacetimes, and higher-order contributions to the interaction. Finally, the entanglement analysis may be extended considering different entanglement measures.

\section*{Acknowledgements}
MD expresses his gratitude to Marco Pettini for interesting debates on topics related to the topic of this work. OL is grateful to Alessio Belfiglio, Edward Bruschi and Marco Muccino for comments and discussions. As pecial thanks to Roberto Della Ceca and to the Brera National Institute of Astrophysics for support. OL acknowledges the Istituto Nazionale di
Fisica Nucleare (INFN) through the Iniziativa Specifica
QGSKY.

\bibliography{bibliography}

\appendix

\section{Mode dynamics in the scalar and vector sectors}{\label{app_modedyamics}}

For the sake of completeness, we report also the equations of motion for the scalar and vector modes. Nonetheless, we remark that, for massless gravitons, these modes do not propagate any physical information. Their dynamics is thus trivial, constrained by the gauge fixing as discussed above.

Firstly, consider Eq. \eqref{eq_EOM-h00-massive} for $h_{00}$. Defining $\tilde{A}_{\bf k}\equiv aA_{\bf k}$, Eq. \eqref{eq_EOM-h00-massive} yields the following equation for $\tilde{A}_{\bf k}$,
\begin{equation}
\tilde{A}_{\bf k}''+\left[k^2-\frac{\nu^2-\frac{1}{4}}{\eta^2}\right]\tilde{A}_{\bf k}=0,\quad \text{with}\quad \nu=\frac32.
\label{eq_EOM-Atilde}
\end{equation}
%
\begin{comment}
Assuming Bunch-Davies initial condition \cite{Bunch:1978yq} in Eq. \eqref{eq_EOM-Atilde}, for $A_{\bf k}$ we find
%
\begin{equation}
A_{\bf k}(\eta)=\frac{1}{a(\eta)}\frac{\sqrt\pi}{2}e^{i\frac{\pi}{2}\left(\nu+\frac{1}{2}\right)}\sqrt{-\eta}H^{(1)}_{\nu}(-k\eta),
\label{eq_A-sol-massive}
\end{equation}
%
where $H^{(1)}_{\nu}(-k\eta)$ is the Hankel function of the first kind of order $\nu$ with argument $-k\eta$.
\end{comment}
Then, consider Eq. \eqref{eq_EOM-h0i-massive} for $h_{0i}$. As $h_{0i}$ contains both scalar and vector modes, Eq. \eqref{eq_EOM-h0i-massive} yields two different equations for $B_{\bf k}$ and $V_{\bf k}$:
\begin{subequations}
\begin{align}
&B_{\bf k}''+\left(k^2-\frac{\nu^2-\frac{1}{4}}{\eta^2}\right)B_{\bf k}=-2\mathcal{H}A_{\bf k},
\label{eq_EOM-B}\\
%\end{equation}
%\begin{equation}
&V_{\bf k}^{(\lambda)}{''}+\left(k^2-\frac{\nu^2-\frac{1}{4}}{\eta^2}\right)V_{\bf k}^{(\lambda)}=0
\label{eq_EOM-V}.
\end{align}
\end{subequations}
%
\begin{comment}
Considering Bunch-Davies initial conditions, Eq. \eqref{eq_EOM-V} gives
%
\begin{equation}
V_{\bf k}^{(\lambda)}(\eta)=\frac{\sqrt\pi}{2}e^{i\frac{\pi}{2}\left(\nu+\frac{1}{2}\right)}\sqrt{-\eta}H^{(1)}_{\nu}(-k\eta).
\label{eq_V-sol-massive}
\end{equation}
%
For $B_{\bf k}$, Bunch-Davies initial conditions cannot be imposed, as the equation has a source term. However, we notice that expanding in Fourier modes Eq. \eqref{eq_transv-explicit} yields
%
\begin{equation}
A_k'+2\mathcal{H}A_k=-k^2B_k,
\label{eq_0-transv-Fourier}
\end{equation}
%
that sets a constraint allowing us to compute $B_{\bf k}$ without solving Eq. \eqref{eq_EOM-B}. Consequently, we obtain
%
\begin{equation}
B_{\bf k}(\eta)\!=\!\frac{\sqrt{\pi}}{2ak}e^{i\frac{\pi}{2}(\nu+\frac{1}{2})}\!\sqrt{\!-\eta}\!\left[\!H_{\nu-1}^{(1)}(-k\eta)\!+\!\frac{\nu\!+\!1/2}{k\eta}H_\nu^{(1)}(-k\eta)\!\right],
\label{eq_B-sol-massive}
\end{equation}
%
that also satisfies Eq. \eqref{eq_EOM-B}.
\end{comment}
Finally, consider Eq. \eqref{eq_EOM-hij-massive}. As $h_{ij}$ contains scalar, vector and tensor modes, Eq. \eqref{eq_EOM-hij-massive} yields an equation for each sector. In particular, we have already found and solved the equation for the tensor sector, namely Eq. (\ref{eq_tensor-sector-hij}), while the other two are
\begin{subequations}
\begin{align}
%\begin{split}
&\delta_{ij}\psi_{\bf k}''-k_ik_jE_{\bf k}''-2\mathcal{H}\left(\delta_{ij}\psi_{\bf k}'-k_ik_jE_{\bf k}'\right)\notag\\&\quad +\left[k^2-\frac{\nu^2-\frac{1}{4}}{\eta^2}\right]\left(\delta_{ij}\psi_{\bf k}-k_ik_jE_{\bf k}\right)\notag\\&\quad \quad =2\mathcal{H}^2\delta_{ij}A_{\bf k}+4\mathcal{H}k_ik_jB_{\bf k}&&(\text{scalar sector}),
%\end{split}
\label{eq_scalar sector-hij}\\
%\end{equation}
%\begin{equation}
%\begin{split}
&F_{\bf k}^{(\lambda)}{''}\!\!-2\mathcal{H}F_{\bf k}^{(\lambda)}{'}\!+\!\left[k^2-\frac{\nu^2-\frac{1}{4}}{\eta^2}\right]F_{\bf k}^{(\lambda)}\notag\\&\quad =-2\mathcal{H}V_{\bf k}^{(\lambda)}&& (\text{vector sector}).
%\end{split}
\label{eq_vector-sector-hij}
\end{align}
\end{subequations}
In addition, Eq. \eqref{eq_scalar sector-hij} can be further separated into an equation for $\psi_{\bf k}$ and another for $E_{\bf k}$, respectively
\begin{subequations}
\begin{align}
&\psi_{\bf k}''-2\mathcal{H}\psi_{\bf k}'+\left(k^2-\frac{\nu^2-\frac{1}{4}}{\eta^2}\right)\psi_{\bf k}=2\mathcal{H}^2A_{\bf k},
\label{eq_EOM-psi}\\
%\end{equation}
%\begin{equation}
&E_{\bf k}''-2\mathcal{H}E_{\bf k}'+\left(k^2-\frac{\nu^2-\frac{1}{4}}{\eta^2}\right)E_{\bf k}=-4\mathcal{H}B_{\bf k}.
\label{eq_EOM-E}
\end{align}
\end{subequations}

\section{Sum over the helicities of the squared amplitude}{\label{app_projectors-momenta}}

Given the amplitude of Eq. \eqref{eq_amplitude-curved-massless}, we obtain
\begin{widetext}
\begin{equation}
\begin{split}
\sum_{\lambda=\pm 2}\!\!\big|\!\!\braket{\hat{\phi}({\bf q_1})\hat{\phi}({\bf q_2})|\hat{S}|\hat{h}({\bf p},\lambda)}\!\!\big|^2\!\!\!&=\!\!\frac{4}{\overline{M}^2_{\text{Pl}}}\frac{V}{(2\pi)^6}\delta^{(3)}\!\left({\bf p}\!-\!{\bf q_1}\!-\!{\bf q_2}\right)\!\left|\mathcal{I}_0({\bf q_1},{\bf q_2},{\bf p})\right|^2\!\!\left[\sum_{\lambda=\pm 2}\epsilon_{ij}^{(\lambda)}({\bf p})\epsilon_{lm}^{(\lambda)*}({\bf p})\right]\!\!(q_1)_i(q_1)_l(q_2)_j(q_2)_m\\&=\frac{4}{\overline{M}^2_{\text{Pl}}}\frac{V}{(2\pi)^6}\delta^{(3)}\left({\bf p}-{\bf q_1}-{\bf q_2}\right)\left|\mathcal{I}_{0}({\bf q_1},{\bf q_2},{\bf p})\right|^2P_{ij,lm}({\bf p})(q_1)_i(q_1)_l(q_2)_j(q_2)_m,
\end{split}
\label{eq_amplitude-curved-massless-intermediate}
\end{equation}
\end{widetext}
where we use Eq. \eqref{eq_propr-compl-pol-tens} and the projector is defined in Eq. \eqref{eq_projector-massive}. In particular, for $P_{ij,lm}({\bf p})(q_1)_i(q_1)_l(q_2)_j(q_2)_m$, we find
\begin{widetext}
\begin{equation}
\begin{split}
&P_{ij,lm}({\bf p})(q_1)_i(q_1)_l(q_2)_j(q_2)_m=\frac{1}{2}\Pi_{il}({\bf p})\Pi_{jm}({\bf p})(q_1)_i(q_1)_l(q_2)_j(q_2)_m=\frac{1}{2}\left[q_1^2-\frac{({\bf q_1}\cdot{\bf p})^2}{p^2}\right]\left[q_2^2-\frac{({\bf q_2}\cdot{\bf p})^2}{p^2}\right]\\&=\frac{1}{2p^4}\left[p^2q_1^2-(q_1^2+{\bf q_1}\cdot{\bf q_2})^2\right]\left[p^2q_2^2-(q_2^2+{\bf q_1}\cdot{\bf q_2})^2\right]=\frac{1}{2p^4}\left[q_1^2q_2^2-({\bf q_1}\cdot{\bf q_2})^2\right]^2=\frac{q_1^4q_2^4(1-x^2)^2}{2p^4},
\end{split}
\label{eq_projector+momenta}
\end{equation}
\end{widetext}
where we use ${\bf p}={\bf q_1}+{\bf q_2}$ and $p^2=q_1^2+q_2^2+2{\bf q_1}\cdot{\bf q_2}$, with ${\bf q_1}\cdot{\bf q_2}=q_1q_2\cos(\theta)=q_1q_2x$.

Then, it is immediate to see that Eq. \eqref{eq_amplitude-curved-massless-intermediate} yields Eq. \eqref{eq_summed-amplitude2-curved-massless}.

\section{Number density of produced particles}{\label{app_Imassless}}

In this section, we compute $\mathcal{I}_0(q_1,q_2,x)$ when using the complete mode solutions for both the scalars and the graviton. Starting from Eq. (\ref{eq_Imassless}), we get
\begin{widetext}
\begin{equation}
\begin{split}
\mathcal{I}_0(q_1,q_2,x)&=\int_{\eta_*}^{\eta_\text{f}}\frac{e^{-ip\eta}}{\sqrt{2p}}\left(\frac{p\eta-i}{p\eta}\right)\left(-\frac{1}{H_I\eta}\right)\frac{H_I}{\sqrt{2q_1^3}}\left(-i-q_1\eta\right)e^{iq_1\eta}\frac{H_I}{\sqrt{2q_1^3}}\left(-i-q_2\eta\right)e^{iq_2\eta}d\eta=\\
&=\frac{H_I}{(2q_1q_2p)^{\frac32}}\int_{\eta_*}^{\eta_\text{f}}\frac{(i-p\eta)(i+q_1\eta)(i+q_2\eta)}{\eta^2}e^{-i(p-q_1-q_2)\eta}d\eta=\\
&=\frac{H_I}{(2q_1q_2p)^{\frac32}}\int_{k_*}^{k_\text{f}}\left(i+\frac{p}{k}\right)\left(i-\frac{q_1}{k}\right)\left(i-\frac{q_2}{k}\right)e^{i\left(\frac{p-q_1-q_2}{k}\right)}dk.
\end{split}
\end{equation}
\end{widetext}
We define
\begin{equation}
\mathcal{B}(q_1,q_2,p)\equiv \int_{k_*}^{k_\text{f}}\left(i+\frac{p}{k}\right)\left(i-\frac{q_1}{k}\right)\left(i-\frac{q_2}{k}\right)e^{i\left(\frac{p-q_1-q_2}{k}\right)}dk,
\end{equation}
yielding Eq. (\ref{eq_I0-entanglement}). We obtain
\begin{widetext}
\begin{equation}
\begin{aligned}
&\mathcal{B}(q_1,q_2,p)=\frac{e^{-\frac{i(k_*+k_{\text{f}})(q_1+q_2)}{k_*k_{\text{f}}}}}{k_* k_{\text{f}}(-p+q_1+q_2)^2}\Bigg\{
i e^{\frac{i(k_{\text{f}}p+k_*(q_1+q_2))}{k_*k_{\text{f}}}}
k_{\text{f}}
\Bigg[
p q_1 q_2(-p+q_1+q_2)
+k_*^2(-p+q_1+q_2)^2+\\
&+i k_*
\left(
p^2(q_1+q_2)
+q_1q_2(q_1+q_2)
-p(q_1^2+4q_1q_2+q_2^2)
\right)
\Bigg]-i e^{\frac{i(k_*p+k_{\text{f}}(q_1+q_2))}{k_*k_{\text{f}}}}
k_*
\Bigg[
p q_1 q_2(-p+q_1+q_2)
+\\
&+k_{\text{f}}^2(-p+q_1+q_2)^2+i k_{\text{f}}
\left(
p^2(q_1+q_2)
+q_1q_2(q_1+q_2)
-p(q_1^2+4q_1q_2+q_2^2)
\right)
\Bigg]
\Bigg\}.
\end{aligned}
\end{equation}
\end{widetext}

\end{document}